\documentclass{article}

\usepackage{PRIMEarxiv}

\usepackage[utf8]{inputenc} 
\usepackage[T1]{fontenc}    
\usepackage{hyperref}       
\usepackage{url}            
\usepackage{booktabs}       
\usepackage{amsfonts}       
\usepackage{nicefrac}       
\usepackage{microtype}      
\usepackage{lipsum}
\usepackage{fancyhdr}       
\usepackage{graphicx}       
\usepackage{multirow}
\graphicspath{{media/}}     

\usepackage{amsmath}
\usepackage{amssymb}
\usepackage{bm,bbm}
\usepackage{xcolor}
\usepackage{natbib} 

\hypersetup{citecolor={blue}, urlcolor={blue} , linkcolor={blue}, colorlinks={true}}

\title{Microstructural and Mechanistic Insights into the Tension-Compression Asymmetry of Rapidly Solidified Fe-Cr Alloys: A Phase Field and Strain Gradient Plasticity Study
}

\author{
  Namit Pai, Indradev Samajdar, Anirban Patra*  \\
  Department of Metallurgical Engineering and Materials Science
  \\
  Indian Institute of Technology Bombay \\
  Mumbai, 400076, India \\
  \texttt{*Corresponding author:anirbanpatra@iitb.ac.in} \\
}

\begin{document}
\maketitle

\begin{abstract}
Rapid solidification in Additively Manufactured (AM) metallic materials results in the development of significant microscale internal stresses, which are attributed to the printing induced dislocation substructures. The resulting backstress due to the Geometrically Necessary Dislocations (GNDs) is responsible for the observed Tension-Compression (TC) asymmetry. We propose a combined Phase Field (PF)-Strain Gradient $J_2$ Plasticity (SGP) framework to investigate the TC asymmetry in such microstructures. The proposed PF model is an extension of Kobayashi’s dendritic growth framework, modified to account for the orientation-based anisotropy and multi-grain interaction effects. The SGP model has consideration for anisotropic temperature-dependent elasticity, dislocation strengthening, solid solution strengthening, along with GND-induced directional backstress. This model is employed to predict the solute segregation, dislocation substructure and backstress development during solidification and the post-solidification anisotropic mechanical properties in terms of the TC asymmetry of rapidly solidified Fe-Cr alloys. It is observed that higher thermal gradients (and hence, cooling rates) lead to higher magnitudes of solute segregation, GND density, and backstress. This also correlates with a corresponding increase in the predicted TC asymmetry. The results presented in this study point to the microstructural factors, such as dislocation substructure and solute segregation, and mechanistic factors, such as backstress, which may contribute to the development of TC asymmetry in rapidly solidified microstructures.   
\end{abstract}



\section{Introduction}
\label{sec:intro}
Additive Manufacturing (AM) of metallic systems can be used for fabrication of complex geometries via incremental deposition of the rapidly solidified melt to conform to the desired geometry, thus reducing the need for post-deposition machining of the AM-ed parts \citep{murr2012metal, gao_status_2015, herzog2016additive, debroy2018additive}. However, this manufacturing technique is generally associated with high solidification (and cooling) rates ($>10^{3}\ \mbox{K}/\mbox{s}$) \citep{farshidianfar_effect_2016, prasad_towards_2020} and severe thermal gradients ($10^{4}-10^{6}\ \mbox{K}/\mbox{m}$) \citep{scipioni_bertoli_stability_2019, pinomaa_significance_2020}, which may result in non-equilibrium microstructures with significant defect densities \citep{gao_status_2015, du2020effects, goel2020built, sanaei_defects_2021}. These defects range from the nanoscale, such as dislocation entrapments at the chemical cell walls \citep{chen_microscale_2019, zhang_modeling_2022}, to the macroscale, such as fusion defects and porosities \citep{beretta_comparison_2017, sanaei_defects_2021}. Further, this may also result in an incomplete solute partitioning at the solid-liquid interface, especially in alloys containing a significant concentration of solute elements. This phenomenon of solute trapping leads to the segregation of solute atoms as well as precipitates along the cell walls and geometrically necessary boundaries, thus promoting dislocation pinning \citep{bertsch2020origin, voisin2021new} and increasing the material's strength, without sacrificing the tensile ductility \citep{liu2018dislocation, wang_additively_2018}. Such a superior strength-ductility combination has been attributed to the development of heterogeneous microstructures in AM materials \citep{wang_additively_2018}.

Understanding these dislocation substructures is hence essential for developing AM microstructures exhibiting superior mechanical properties. Various mechanisms have been identified for the development of such heterogeneous dislocation substructures \citep{saeidi2015hardened, liu2018dislocation, yoo2018identifying, birnbaum2019intrinsic, bertsch2020origin, zhang_additively_2021, voisin2021new}. For example, \cite{saeidi2015hardened} reported the accommodation of distortion induced by the solute enriched sub-grain and inter-dendritic boundaries via dislocation networks and substructures in AM austenitic Stainless Steel (SS 316L). The formation of cell boundaries in AM Inconel 718 has also been attributed to a combination of solute supersaturation at the solid-liquid interface and subsequent dislocation generation to accommodate the lattice incompatibilities \citep{yoo2018identifying}. The Geometrically Necessary Dislocations (GNDs), which originate during solidification for accommodating the inter-dendritic misorientations, further act as a scaffold for the mobile dislocations \citep{bertsch2020origin}. \cite{ramirez2011novel} attributed the presence of fine cuprite precipitates, along the inter-dendritic as well as grain boundary regions, for the development of complex precipitate-dislocation networks in AM copper components. An alternative mechanism suggests that the thermal distortion occurring due to rapid solidification (and cooling) results in the formation of cellular structures, which then act as sinks for the solute and precipitate atoms in AM SS 316L \citep{birnbaum2019intrinsic}. In addition, the boundary constraints surrounding the molten pool too play a role in deciding the dislocation substructures during rapid solidification \citep{bertsch2020origin}. In summary, the printing induced stresses, solute atoms, precipitates and the local inter-dendritic misorientation play an important role in the formation of the observed dislocation cells during rapid solidification.  

In terms of the mechanical properties, rapid solidification leads to the development of internal (residual) stresses, which govern the ensuing deformation behavior of AM-ed parts \citep{wu2014experimental, brown_situ_2017, brown_using_2019, chen_microscale_2019}. Briefly, these can be divided into macroscale (bulk or type-I) and microscale (type-II and type-III) residual stresses \citep{verlinden2007thermo, lodh2017relating}. The former are mainly attributed to the specimen geometry, AM process parameters, scanning strategy, clamping constraints, etc. \citep{ding2011thermo, wu2014experimental, kumar2023thermal}, and can be reduced by optimizing the process parameters or relaxed by subsequent annealing. The latter can be further classified into two categories: intergranular (type-II) and intragranular (type-III) internal stresses \citep{chen_microscale_2019, zhang_modeling_2022}. The intergranular internal stresses are a consequence of the compatibility conditions arising at the grain boundaries, and self-equilibrate over the length scale of a few grains \citep{zhang_modeling_2022}. The intragranular stresses are primarily due to the printing-induced dislocation substructures, specifically the GND components of these dislocation clusters \citep{zhang_modeling_2022}. The heterogeneous distribution of GNDs thus gives rise to backstresses, and are hypothesized to be responsible for the anisotropic mechanical properties, including Tension-Compression (TC) asymmetry in rapidly solidified alloys \citep{chen_microscale_2019, zhang_modeling_2022, wang_anisotropic_2023}. Further, the as-printed dislocation substructures act as forest obstacles to dislocation slip, and influence the associated mechanical properties \citep{wang_additively_2018, bean_heterogeneous_2022}. For example, AM austenitic steels (SS 316L) exhibit a significantly higher yield strength and total elongation to failure, in comparison to the as-cast and wrought counterparts \citep{sun2018simultaneously, wang_additively_2018}. The substructure development is generally dependent on the local thermal history, which is governed by the imposed thermal gradient and the pulling velocity during AM \citep{pinomaa_significance_2020, lindroos_dislocation_2022}.

Much work has been done on modeling the mechanical properties of AM metals post-solidification by accounting for the effect of microstructure and residual stresses  \citep{kapoor2018incorporating, chen_microscale_2019, li_tensile_2019, pokharel_analysis_2019, somlo2021anisotropic, zhang_modeling_2022, wang_anisotropic_2023}. Further, multiscale models coupled with the phase field method  \citep{azizi_characterizing_2021, hu2022microscale, lindroos_dislocation_2022, pinomaa_multiscale_2022} have been developed for simulating the solidification and grain growth processes as well. Macroscale finite element models have been used to mimic the temperature profiles of various AM processes, for example, Selective Electron Beam Melting (SEBM) and Laser Powder Bed Fusion Process (LPBF), which then served as input to a temperature-dependent phase field model \citep{liu_investigation_2018, liu2020integration, azizi_characterizing_2021}. Moreover, the preferred growth direction and crystallographic orientation based anisotropy were also accounted for by altering the gradient energy coefficients \citep{liu_investigation_2018}. Several studies have utilized mechanics coupled phase field models to analyze the effect of solidification parameters (and consequent thermal profiles) on the microsegregation, Primary and Secondary Dendrite Arm Spacing (PDAS/SDAS), grain nucleation and growth and interface morphologies in rapidly solidified microstructures \citep{farzadi_phase-field_2008, yu_phase_2018, lingda_phase-field_2021}.

These frameworks have also been coupled with crystal plasticity models, to study the microsegregation, grain-level residual stresses and dislocation structure developments during AM \citep{pinomaa_significance_2020, lindroos_dislocation_2022, hu2022microscale, hu2023dislocation}. \cite{pinomaa_significance_2020, lindroos_dislocation_2022} have used a phase field-crystal plasticity model to simulate the microstructure evolution (dislocation structure and solute segregation) during solidification. \cite{hu2022microscale} also utilized such an approach and observed significant plastic deformation in (a few) elongated grains along the laser track, as a result of the imposed thermal and geometric constraints. Homogenization of these plastic strains lead to reduction in the residual stresses in the AM microstructure \citep{hu2022microscale}. Further, \cite{hu2023dislocation} have used the temperature as well as microstructure histories that act as input to a 3D Continuum Dislocation Dynamics (CDD) model to study the local evolution of dislocation structures during a laser-based AM process. \cite{kuna2023framework} have employed a Cellular Automata Finite Element framework to predict the microstructure evolution during LPBF-based solidification of a SS316L alloy. Following this, the microstructure as well as thermal history is then imported into a thermo-mechanical crystal plasticity model, for predicting the residual stress and strain distributions using various processing parameters \citep{kuna2023framework}. However, these models do not account for the strain gradient plasticity induced backstresses, and its influence on the anisotropic mechanical properties. In this regard, \cite{chen_microscale_2019, zhang_modeling_2022} have shown that by phenomenologically introducing a constant value of backstress, representative of the printing-induced residual stresses, in their crystal plasticity model, the experimentally observed TC asymmetry of rapidly solidified microstructures can be predicted. However, their models did not simulate the \textit{in-situ} development of printing induced residual stresses. In the present work, we try to address these gaps in the literature by developing a unified modeling framework for predicting the development of GND-induced backstress in rapidly solidified microstructures and the ensuing anisotropic mechanical properties.

This manuscript proposes a coupled phase field (PF) - strain gradient plasticity (SGP) framework, for simulating microstructure evolution due to dislocation substructures and microsegregation, and predicting the Tension-Compression (TC) asymmetry in rapidly solidified iron-chromium (Fe-Cr) alloys. Essential features of the framework include consideration for solute segregation, anisotropic elastic and plastic deformation, GND and backstress evolution during solidification, multi-grain interaction effects, as well as thermally-induced residual stress development and their accommodation via plastic deformation during solidification. Simulations are performed for a range of solidification conditions, representative of AM conditions, to identify the microstructural and mechanistic origins of TC asymmetry in rapidly solidified Fe-Cr alloys. Finally, an analysis of the modeling framework itself is performed to identify the essential features of the model necessary for predicting the TC asymmetry.

\section{Model Description}
This work uses a combined phase field (PF)-strain gradient plasticity (SGP) modeling approach for studying the factors contributing to the development of microscale residual stresses in rapidly solidified microstructures and their effect on the mechanical properties of iron-chromium (Fe-Cr) alloys. The phase field model is used to simulate solidification in idealized 2D polycrystalline microstructures, along with consideration for the solute (Cr) segregation during solidification. Further, the phase field model is coupled with a $J_2$ SGP model that allows consideration for the directional, GND-induced backstresses in the solidified phase. These backstresses are expected to be induced while accommodating the thermal Eigen strains during solidification and cooling and also the residual stresses due to Cr segregation during solidification. Solid solution strengthening due to the Cr atoms is also considered. The solidified microstructures are then virtually deformed using the SGP model to study the anisotropy of mechanical properties, more specifically, the Tension-Compression (TC) asymmetry along the build and transverse directions in these rapidly solidified microstructures.

The PF model and the SGP model are loosely coupled in the sense that the dissipation of work due to plastic deformation in the solidified phase is not considered in the free energy minimization of the PF model. Similar assumption has also been made in prior studies \citep{pinomaa_significance_2020, hu2022microscale, lindroos_dislocation_2022}. Further, the use of a $J_2$ SGP model, instead of a strain gradient crystal plasticity model, is necessitated by the feasibility of running these coupled phase field-plasticity simulations, which are computationally intensive. We note that there are more advanced strain gradient crystal plasticity models \citep{evers_non-local_2004} and thermo-mechanical crystal plasticity models \citep{chen_microscale_2019, pokharel_analysis_2019} in the literature, for example. However, coupling these crystal plasticity models with the PF models led to a significant increase in the computational costs and meaningful results could not be realized for the problem of interest here. This is also the reason that idealized 2D microstructures have been simulated here, instead of 3D polycrystalline microstructures. The following sections provide details on the development and implementation of our coupled PF-SGP model.

\subsection{Phase Field Model}
\label{phase_field_sec}
The phase field method is a versatile technique for simulating the mesoscale microstructure evolution during solidification, grain growth and phase transformation phenomena that models the evolution of the order parameter, representative of the transformation from one phase/state to another, by minimizing the free energy of the system \citep{kobayashi_modeling_1993, yeon2001phase, militzer2011phase, permann2016order, acharya_prediction_2017, biswas_solidification_2022}. In the present phase field model, the microstructural features are modeled in terms of a set of continuous variables that evolve in the spatial as well as temporal coordinates: 
\begin{enumerate}
    \item	The non-conserved order parameters, $\phi_{i}$ ($0 \leq \phi_{i} \leq 1$ and $i \in 1,2,….N$), which describe the microstructure in terms of the phase (i.e., liquid, $\phi_{i} = 0$, or solid, $\phi_{i} = 1$) for the $N$ grains, with unique crystal orientations, considered in the simulation.
    \item The conserved variable, $c_{Cr}$, representing the concentration of the solute (Cr) in the Fe system.
\end{enumerate}

\subsubsection{Evolution of the Order Parameter, \  $\phi_i$}
We adopt the formulation given by \cite{kobayashi_modeling_1993} to describe the evolution of our order parameter/phase field variable, $\phi_i$, and further modify it to account for the interaction between $i^{th}$ and $j^{th}$ grain of the solid phase \citep{pinomaa_phase_2020, biswas_solidification_2022}. This is given as:
\begin{equation}
    \label{phi_evolution}
    \tau \frac{\partial \phi_i}{\partial t}=-\frac{\partial}{\partial x}\left(W W^{\prime} \frac{\partial \phi_i}{\partial y}\right)+\frac{\partial}{\partial y}\left(W W^{\prime} \frac{\partial \phi_i}{\partial x}\right)+ \nabla \cdot\left(W^2 \nabla \phi_i\right)+\phi_i\left(1-\phi_i\right)\left(\phi_i-\frac{1}{2}+m\right)+ 2\phi_i \sum_i \sum_j \gamma_{s_{i j}} \phi_j^2
\end{equation}
where, $\tau$ and $W$ represent the interface relaxation time and the anisotropic interface layer width, respectively. $W^{'}$ denotes the first derivative of $W$ with respect to $\psi$, the angle subtended between the interface normal and the global $X$ direction. Mathematically, this can be written as \citep{kobayashi_modeling_1993}:
\begin{equation}
    \psi=\tan ^{-1}\left(\frac{\partial \phi_i /\partial y}{\partial \phi_i /\partial x}\right)
\end{equation}
Further, ${\gamma_{s}}_{ij}$ is representative of the interfacial energy, $\gamma_{s}$, between the $i^{th}$ and $j^{th}$ grain \citep{pinomaa_phase_2020, biswas_solidification_2022}. The thermodynamic driving force for microstructure evolution is written in terms of the parameter, $m$, as \citep{kobayashi_modeling_1993}:
\begin{equation}
    m=\frac{\beta}{\pi} \tan ^{-1}\left(\Gamma\left(\theta_e-\theta_r\right)\right)
\end{equation}
where, $\beta$ and $\Gamma$ are material constants and $\theta_e$ is the non-dimensional equilibrium temperature, which has been set to unity. Further, $\theta_r$ is the non-dimensional temperature ranging from $0$ to $1$ and can be written as a function of the local temperature, $\theta$, as:
\begin{equation}
\label{eqn:reduced_temperature}
    \theta_r=\frac{\theta_m-\theta}{\theta_m-\theta_{ref}}
\end{equation}
where, $\theta_m$ and $\theta_{ref}$ denote the melting point and reference temperature, respectively. The local temperature in absolute units, $\theta$, which evolves spatially and temporally, has been non-dimensionalized here to $\theta_{r}$, such that $\theta_{r}=1$ indicates a fully solidified melt and $\theta_{r}=0$ indicates a fully liquid region \citep{kobayashi_modeling_1993}. The thermodynamic driving force is directly proportional to the degree of undercooling, $\theta_{e}-\theta_{r}$, i.e., higher undercooling results in a rapid evolution of $\phi_{i}$ fields across the liquid region. The evolution of the local temperature $\theta$ is governed by the imposed thermal gradient $G$ during directional solidification. This has been discussed later in Section \ref{coupled_PF_SGP}. 

Using the law of conservation of enthalpy, we can write \citep{kobayashi_modeling_1993, acharya_prediction_2017}:
\begin{equation}
    \frac{\partial \theta_r}{\partial t}=\nabla^2 \theta_r+K \sum_{i=1}^N \frac{\partial \phi_i}{\partial t}
\end{equation}
where, $K$ is the dimensionless latent heat, which varies directly with the latent heat, $L$, and inversely with the specific heat, $C_{p}$, i.e., $K=f(L/C_p )$ \citep{kobayashi_modeling_1993, acharya_prediction_2017}. The moving heat source term, $K \sum_{i=1}^N \partial \phi_i / \partial t$, is only present at the interface where the phase field variable, $\phi_i$, transitions from zero to unity and is zero elsewhere \citep{kobayashi_modeling_1993}.

Prior studies have shown that the anisotropy in thermal conductivity, elastic modulus and surface energy of a grain can inherently determine its preferred growth direction \citep{lee_factors_1997}. More importantly, these properties influence the favored growth direction by minimizing the strain and the crystal/liquid interface energy. In the present work, the interface layer width, $W$, has been modified to take into account the inherent anisotropy as a function of the growth direction \citep{kobayashi_modeling_1993, acharya_prediction_2017}, as:
\begin{equation}
\label{eqn:anisotropic_interfacial_width}
    W=\bar{W}\left(1 + \varsigma_i \cos (\Omega \psi)\right)
\end{equation}
where, $\bar{W}$ is the mean value of the interface width $W$, $\varsigma_{i}$ is the anisotropy strength of the $i^{th}$ grain and $\Omega$ is the anisotropy mode number. \cite{liu_investigation_2018} accounted for anisotropic grain growth by altering the gradient energy coefficients, such that the grains with $<001>$ axis oriented closer to the thermal gradient have a competitive advantage during grain growth. Accordingly, the anisotropy strength, $\varsigma_{i}$, has been rewritten as a function of the crystallographic orientation of the grain $i$ as \citep{liu_investigation_2018}:
\begin{equation}
    \label{anisotropic_grain_growth}
   \varsigma_i=\bar{\varsigma}\left|\cos \angle\left(<001>_{\text {axis }}, \nabla T\right)\right|
\end{equation}
where, $\bar{\varsigma}$ is a numerical constant, signifying the average anisotropic strength. Physically, the above equation indicates that grains having $<001>$ axis oriented closer to the direction of thermal gradient have a higher $\varsigma_{i}$, thus exhibiting a stronger preference during the grain growth.

The interface between two grains $i$ and $j$ is given by a smooth variation of the order parameter $\phi_i$ from $1$ to $0$ and vice-versa for $\phi_j$ (cf. Equation \ref{phi_evolution}). This is in contrast to the sharp interface models, that inherently assume that the grain/phase interfaces are infinitely sharp, i.e., the properties exhibit a sudden discontinuity at these interfaces \citep{liu2004grain,detor2007grain}. On the other hand, the diffuse interface models are defined by a set of variables that are continuous functions of space and time \citep{cha_phase_2002, kim_thermodynamic_2016,zhou2020numerical}.

\subsubsection{Evolution of Cr Concentration, \ $c_{Cr}$}
Chemically, the system is considered to be a single phase, i.e., the chemical activity of all grains in the system is denoted by a single (continuous) conserved parameter representing the Cr concentration, $c_{Cr}$, as opposed to the non-conserved order parameters, $\phi_i$, which are considered for individual grains \citep{pinomaa_phase_2020}.

We use a regular solution model to predict the evolution of the conserved variable, $c_{Cr}$. In this context, an idealized pseudo binary Fe-Cr alloy has been assumed, with the Cr solute randomly distributed in the Fe solvent. This is motivated from prior studies \citep{bertsch2020origin, voisin2021new} that have reported Cr as the primary segregating element in AM stainless steels. Accordingly, the chemical free energy of the system can be written as:
\begin{equation}
    \label{regular_solution_free_energy}
    F_{\text{chem}}=G_{\text{liq }}^0\left(1-\phi_i\right)+\phi_i (G_{F e}^0 c_{F e}+G_{C r}^0 c_{C r}+ G_m^E + \\R T\left(c_{F e} \ln \left(c_{F e}\right) + c_{C r} \ln \left(c_{C r}\right)\right))
\end{equation} 
where, $c_{Cr}$ and $c_{Fe}$ represents the concentration of Cr (solute) and Fe (solvent) atoms, respectively. The terms accounting for the free energy contributions from pure components ($G_{Cr}^{0}$, $G_{Fe}^{0}$), liquid melt ($G_{\text {liq}}^0$) and the heat of mixing ($G_{m}^{E}$) are obtained from thermodynamic databases of equilibrium phase diagrams \citep{dinsdale_sgte_1991, miettinen_thermodynamic_1999, koyama_phase-field_2004, kim_establishment_2021}. Here, $R$ is the gas constant. Further, by substituting $c_{Fe}=1-c_{Cr}$, the individual terms in Equation \ref{regular_solution_free_energy} can be written as:
\begin{equation}
    G_{C r}^0=-8856.9+157.48 \theta-26.908 \theta \ln (\theta)+0.00189 \theta^2-1.477 \times 10^{-6} \theta^3+\frac{139250}{\theta} \mbox{J}/\mbox{mol}
\end{equation}
\begin{equation}
    G_{F e}^0=-236.7+132.416 \theta-24.66 \theta \ln (\theta)-0.003757 \theta^2-5.86 \times 10^{-8} \theta^3+\frac{77358.5}{\theta} \mbox{J}/\mbox{mol}
\end{equation}
\begin{equation}
    G_{\text {liq}}^0=-180383.83+291302 \theta-46.01 \theta \ln (\theta) \mbox{J}/\mbox{mol}
\end{equation}
\begin{equation}
    \label{free_energy_mix}
    G_m^E=\left((10833-7.477)-1410\left(1-2c_{Cr}\right)\right) (1-c_{Cr}) c_{Cr} \mbox{J}/\mbox{mol}
\end{equation}
The solute diffusivity in the solid is expected to be significantly lower than that in the liquid \citep{lindroos_dislocation_2022}. Moreover, increasing the interface thickness results in a magnification of the solute trapping effect \citep{echebarria2004quantitative, ghosh2017primary}. In order to restore the local equilibrium at the interface, an additional solute back current is introduced that pushes the solute out of the solidified region. This back current is referred to as the anti-trapping current \citep{plapp_unified_2011, biswas_solidification_2022}. The modified Cahn-Hilliard equations governing the evolution of solute concentration can then be written as \citep{pinomaa_significance_2020, lindroos_dislocation_2022, pinomaa_multiscale_2022}: 
\begin{equation}
\label{eqn:c_evol}
    \frac{\partial c_{Cr}}{\partial t}=\nabla \cdot \left[ M \nabla \frac{\partial F_{\text {chem }}}{\partial c_{Cr}}+  W_0  \left(1-k_e\right) \frac{c_{Cr}}{c_{Cr}^{eq}} \sum_{i} a_t^{\prime} \frac{\partial \phi_i}{\partial t} \frac{\nabla \phi_i}{|\nabla \phi_i|} \right]
\end{equation}
where, $c_{Cr}^{eq}$ can be derived using the equilibrium partition coefficient, $k_e$, as $c_{Cr}^{eq}=\frac{1+k_e-(1-k_e)h}{2}$, where $h=N-1+\Sigma_i \phi_i$ \citep{pinomaa_phase_2020}. Further, $W_0$ is the magnitude of interface width and $a_{t}^{'}$ represents the modified anti-trapping coefficient, given by \citep{lindroos_dislocation_2022, pinomaa_significance_2020, pinomaa_multiscale_2022}: 
\begin{equation}
    a_t^{\prime}=\frac{1}{2 \sqrt{2}}\left(1-A\left(1-\phi_i^2\right)\right)
\end{equation}
Here, $A$ represents the solute trapping parameter. Further, using the approach given by \cite{koyama_phase-field_2004}, the interface mobility, $M$, is written as:
\begin{equation}
    \label{interface_mobility_chem}
    M=\left[c_{Cr}(1-c_{Cr}) D_{F e}^*+(1-c_{Cr})^2 D_{C r}^*\right]\left(\frac{c}{R \theta}\right)
\end{equation}
where, $D_{Fe}^{*}$ and $D_{Cr}^{*}$ denote the self-diffusion constant of the solvent and the solute, respectively.

We have used the open-source finite element library, Multiphysics Object Oriented Simulation Environment (MOOSE) \citep{permann_moose_2020} for the phase field and plasticity simulations. A Kobayashi formulation-based single crystal dendritic growth framework \citep{kobayashi_modeling_1993} is already implemented in MOOSE \citep{zhou2020numerical}. The existing PF models in MOOSE have been modified from their existing form to those given in Equations \ref{phi_evolution}-\ref{anisotropic_grain_growth} and Equations \ref{regular_solution_free_energy}-\ref{interface_mobility_chem}, respectively. Specifically, we have modified the existing formulation to account for the multigrain interaction effects ($5^{th}$ term in Equation \ref{phi_evolution}) and the crystallographic orientation-based anisotropy during grain growth (Equation \ref{anisotropic_grain_growth}). In addition, we have modified the existing Cahn-Hilliard model to account for the regular solution free energy given in Equation \ref{regular_solution_free_energy}.

\subsubsection{Parameters for the Phase Field Model}

The model parameters used to study the evolution of the above-mentioned phase field variables are summarized in Table \ref{table:phase_field_calib}. The constants $\beta$ and $\Gamma$, which define the double-well potential curve, the numerical constant $\tau$, governing the interface attachment timescale and the dimensionless latent heat, $L$ have been taken directly from Kobayashi's work \citep{kobayashi_modeling_1993}. The average anisotropic strength, $\bar{\varepsilon}$, and the mode number, $\Omega$, were obtained from \cite{zhou2020numerical}.The interfacial energy constant was taken from \cite{pinomaa_phase_2020} .The equilibrium Cr concentration $c_{Cr}^{0}$ and melting temperature and $\theta_m$ were taken from \cite{mcguire2008stainless}, while the parameters $k_e$ and $A$ were adopted from \cite{lindroos_dislocation_2022}, which also modeled a similar Fe-Cr alloy system. The average interface width, $\bar{W}$, was selected such that sufficient number of mesh elements existed within the interface, to achieve a continuous transition from liquid to the solid phase. The self-diffusion constants for Fe and Cr were taken from \cite{koyama_phase-field_2004}.

\begin{table}[!htbp]
   \begin{center}
     \caption{Phase field model parameters for a representative pseudo binary Fe-Cr alloy.}
     \label{table:phase_field_calib}
     \footnotesize
     \begin{tabular}{c c c}
      \hline
         Parameter & Description & Value \\ \hline
         $\beta, \Gamma$ & Constants governing the slope of double well potential \citep{kobayashi_modeling_1993}  & $0.9, 10$ \\
         $\tau$ & Numerical constant \citep{kobayashi_modeling_1993} & $0.0003$ \\ 
         ${\gamma_s}_{ij} \forall\ i,j$ & Interfacial energy constant \citep{pinomaa_phase_2020} & $20$ \\ 
         $\bar{\varepsilon}$ & Average anisotropic strength \citep{kobayashi_modeling_1993,zhou2020numerical} & $0.04$ \\ 
         $\Omega$ & Mode number & $4$ \\ 
         $\bar{W}$ & Average interface width ($\mu \mbox{m}$) & $0.14$ \\ 
         $K$ & Dimensionless latent heat \citep{kobayashi_modeling_1993} & $1.20$ \\ 
         $\theta_m$ & Melting temperature (K) \citep{mcguire2008stainless} & $1789$ \\ 
         $k_{e}$ &Equilibrium partition coefficient \citep{lindroos_dislocation_2022} & $0.79$ \\ 
         $c_{Cr}^{0}$ & Equilibrium Cr concentration (wt. fraction) \citep{mcguire2008stainless} & $0.21$ \\ 
         $A$ & Solute trapping parameter \citep{lindroos_dislocation_2022} & $0.88$ \\ 
         $D_{Fe}^{*}$ & Self-diffusion constant for Fe ($\mbox{m}^2/\mbox{s}$) \citep{koyama_phase-field_2004} & $1 \times 10^{-4} \exp \left(-\frac{294 \times 10^3}{R T}\right)$ \\ 
         $D_{Cr}^{*}$ & Self-diffusion constant for Cr ($\mbox{m}^2/\mbox{s}$) \citep{koyama_phase-field_2004} & $2 \times 10^{-5} \exp \left(-\frac{308 \times 10^3}{R T}\right)$ \\ \hline
     \end{tabular}
   \end{center}
\end{table}


\subsection{Strain Gradient Plasticity Model}
\label{sec:SGP_model}
The finite deformation $J_2$ plasticity model is adapted from \cite{patra_modeling_2023}, and is based on the multiplicative decomposition of the deformation gradient, $\bm{F}$, into the elastic, $\boldsymbol{F}^{e}$, plastic, $\boldsymbol{F}^{p}$, and thermal, $\boldsymbol{F}^{\theta}$, parts \citep{musinski_eigenstrain_2015, pokharel_analysis_2019}, i.e.,
\begin{equation}
    \label{def_grad}
    \boldsymbol{F}=\boldsymbol{F}^{e} \cdot \boldsymbol{F}^{p} \cdot \boldsymbol{F}^{\theta}
\end{equation}
Here, $\boldsymbol{F}^{e}$ accounts for the elastic deformation, $\boldsymbol{F}^{p}$ accounts for the shear due to plastic deformation, and $\boldsymbol{F}^{\theta}$ accounts for the Eigen strain due to thermal expansion/contraction (see Figure \ref{fig:figure_02}). Assuming isotropic thermal expansion/contraction, $\boldsymbol{F}^{\theta}$ can be written as \citep{musinski_eigenstrain_2015}:
\begin{equation}
    \label{F_theta}
    \boldsymbol{F}^{\boldsymbol{\theta}}=\sqrt{1+2 \alpha^\theta \Delta \theta}\  \boldsymbol{I}
\end{equation}
where, $\alpha^{\theta}$ is the thermal expansion coefficient, which is a function of the temperature, ${\theta}$, i.e., $\Delta \theta =\theta_{ref}-\theta$ represents the deviation from the stress-free temperature, $\theta_{ref}$, and $\boldsymbol{I}$ is the second rank identity tensor. The resulting ‘stress-free’ Eigen strains can thus be written as:
\begin{equation}
    \label{E_theta}
    \boldsymbol{E}^{\theta} = \frac{\mathbf{1}}{\mathbf{2}}\left(\boldsymbol{F}^{\theta T} \cdot \boldsymbol{F}^{\theta}-\boldsymbol{I}\right)=\alpha^\theta \Delta \theta\  \boldsymbol{I}
\end{equation}

The evolution of the plastic deformation gradient, $\boldsymbol{F}^{p}$, can be written in terms of the plastic part of the spatial velocity gradient, $\boldsymbol{L}^{p}$, as $\dot{\boldsymbol{F}}^p=\boldsymbol{L}^{p} \cdot \boldsymbol{F}^{p}$. In this $J_2$ plasticity model, $\boldsymbol{L}^{p}$ is simply given by \citep{patra_modeling_2023}:
\begin{equation}
    \label{eqn:equation_19}
    \boldsymbol{L}^{p}=\sqrt{\frac{3}{2}} \dot{\bar \varepsilon}^p \boldsymbol{N}^{p}
\end{equation}
where, $\dot{\bar \varepsilon}^p$ is the effective plastic strain rate and $\boldsymbol{N}^{p}$ is the unit tensor along the direction of plastic flow. $\boldsymbol{N}^{p}$ can be further written in terms of the deviatoric stress, $\boldsymbol{S}$, and the backstress tensor, $\boldsymbol{\chi}$, as \citep{patra_modeling_2023}:
\begin{equation}
    \label{eqn:equation_20}
    \boldsymbol{N}^{p}=\sqrt{\frac{3}{2}} \frac{\boldsymbol{S}-\boldsymbol{\chi}}{\bar{\sigma}^*}
\end{equation}
Here, $\bar{\sigma}^*$ describes the modified effective stress, which is given by: $\bar{\sigma}^*=\sqrt{\frac{3}{2}(\boldsymbol{S} - \boldsymbol{\chi}):(\boldsymbol{S} - \boldsymbol{\chi})}$. Generally speaking, $\boldsymbol{L}^{p}$ can be additively decomposed into a symmetric part, $\overline{\boldsymbol{D}}^{\boldsymbol{p}}=\operatorname{sym}\left[\boldsymbol{F}^p \cdot \boldsymbol{F}^{\boldsymbol{p}^{-1}}\right]$, which is representative of the plastic strain rate and an anti-symmetric part, $\overline{\boldsymbol{W}}^p=\operatorname{ant}\left[\boldsymbol{F}^p \cdot \boldsymbol{F}^{\boldsymbol{p}^{-\mathbf{1}}}\right]$, representative of the plastic spin or the substructure rotation \citep{hashiguchi2020nonlinear, weber1990finite}. The formulations presented in Equations \ref{eqn:equation_19}-\ref{eqn:equation_20} for $\boldsymbol{L}^{p}$ have been adopted from \cite{weber1990finite}, where it was assumed that $\overline{\boldsymbol{D}}^{\boldsymbol{p}}=\operatorname{sym}\left[\dot{\boldsymbol{F}}^{\boldsymbol{p}} \cdot \boldsymbol{F}^{\boldsymbol{p} ^{- 1}}\right]=\sqrt{\frac{3}{2}} \dot{\bar \varepsilon}_p \boldsymbol{N}_{\boldsymbol{p}}$ and $\overline{\boldsymbol{W}}^p=\operatorname{ant}\left[\boldsymbol{F}^p \cdot \boldsymbol{F}^{\boldsymbol{p}^{-1}}\right]=0$, thus neglecting the anti-symmetric part of $\boldsymbol{L}^{p}$ for isotropic plasticity. As a first order approximation, we have also not considered the rotation of the substructure in our anisotropic model. While the effect of plastic spin may be negligible for small deformations, we note that such an approximation may not hold for large plastic strains and appropriate modifications can be made in future work (see \cite{dafalias1985plastic}, for example).

The effective plastic strain rate is modeled using a Kocks-type thermally activated flow rule that can account for temperature- and strain rate-dependent effects \citep{kocks_progress_1975}, i.e.,
\begin{equation}
    \label{eqn:equation_21}
    \dot{\bar \varepsilon}^p=\dot{\bar{\varepsilon}}_0^p \exp \left\{\frac{-\Delta F_g}{k \theta}\left(1-\left(\frac{\bar{\sigma}^*-s_a}{s_t}\right)^p\right)^q\right\}; \bar{\sigma}^* > s_a \\
\end{equation}
where, $\dot{\bar{\varepsilon}}_0^p$ denotes the reference strain rate, $\Delta F_{g}$ is the activation energy for dislocation glide, the athermal slip resistance, $s_a$, accounts for slip resistance due to long range stress fields of dislocation junctions, and the thermal slip resistance, $s_t$, accounts for slip resistance due to short range barriers. $k$ denotes the Boltzmann constant and $\theta$ denotes the absolute temperature, respectively. Further, $p$ and $q$ are parameters governing the shape of the activation enthalpy curve.

The athermal slip resistance, $s_a$, is represented using a typical Taylor-type hardening model as \citep{taylor1934mechanism}:
\begin{equation}
    \label{eqn:athermal_hardening}
    s_a= a \left(\tau_0+k_{I H} G b \sqrt{\rho_{S S D}}\right) 
\end{equation}
where, $\tau_0$ is the threshold resistance, $k_{IH}$ is the (isotropic) Taylor hardening coefficient due to the Statistically Stored Dislocation (SSD) densities, $\rho_{SSD}$, $G$ is the shear modulus, and $b$ is the Burgers vector magnitude. 

The SGP model is by definition isotropic, i.e., it does not take into account the anisotropy induced by the crystallographic orientation or texture of a microstructure. However, this assumption may not be appropriate in the present work, where we intend to simulate solidification in polycrystalline microstructures. As described earlier, the phase field model accounts for the growth of grains with different crystal orientations. Hence, in order to incorporate the effect of anisotropy (due to crystallographic orientation) on the yield surface, the above equation is multiplied by an anisotropy factor, $a$, which is representative of the Taylor factor. Inspired from \cite{patra_modeling_2023}, the factor $a$ is modeled as the inverse of the maximum Schmid factor over all possible slip systems of the crystal structure under consideration, i.e.,
\begin{equation}
  \label{anisotropy_factor}
  a=\frac{1}{\max \left(m^\alpha\right)} ; \alpha=1,2, \ldots, N_s
\end{equation}
where, $m^\alpha$ represents the Schmid factor associated with the $\alpha^{th}$ slip system of the crystal with $N_s$ possible slip systems. Here, $\boldsymbol{s}^\alpha$  and $\boldsymbol{n}^\alpha$, denote the unit slip and slip plane normal directions, respectively. Further, $m_\alpha$ is defined as:
\begin{equation}
    m^\alpha=\frac{\boldsymbol{\sigma}}{\|\boldsymbol{\sigma}\|}: \boldsymbol{s}^\alpha \otimes \boldsymbol{n}^\alpha \\
\end{equation}
where, $\boldsymbol{\sigma}$ is the Cauchy stress. $m^\alpha$ may be considered as the equivalent of the Schmid factor commonly used in crystal plasticity models. Essentially, the factor, $a$, introduces anisotropy along the slip system with the maximum Schmid factor under the assumption of single slip in our $J_2$ plasticity model. Further, the slip system information is only used to compute the anisotropy factor in this otherwise "macro-plasticity" model. We note that such an assumption may not be valid at large plastic strains, when multiple slip is to be expected. 

Anisotropic yield has generally been incorporated in macro-plasticity models using the Hill’s stress potential \citep{hill1948theory} and the associated anisotropic material constants can be derived by testing bulk specimens (sizes greater than a few mm) along different orientations. Given that we do not have much experimental data regarding orientation-dependent yield anisotropy for the idealized Fe-Cr alloy under study, we resort to a crystal plasticity-motivated model, by introducing the anisotropy factor, $a$, which accounts for the orientation-dependence of the slip resistance. The emphasis in the present work is on predicting the anisotropic elastic and plastic deformation in smaller microstructures (less than $100\ \mu m$ size) at the single crystal and polycrystal level using a macro-plasticity model. Predictions from verification simulations are also presented in Appendix \ref{sec:appendix_A}, which demonstrate the model’s ability to predict the crystalline anisotropy-induced deformation at small strains in single crystals. Additionally, it should be noted that a crystal plasticity model naturally accounts for the orientation effect in terms of the resolved shear stress, rather than altering the slip resistance. In our SGP framework, the stress potential is based on the second invariant, $J_{2}$, of the deviatoric stress tensor minus the backstress tensor and resembles a von Mises yield criterion (c.f. Equation \ref{eqn:equation_20}). Without altering this stress potential, we introduce crystalline anisotropy in the model by proposing the anisotropy factor, $a$, which modifies the slip resistance based on the crystal orientation. 

Similar to crystal plasticity models, elastic anisotropy has also been considered in our model as:
\begin{equation}
    \label{eqn:anisotropic_elasticity}
    \boldsymbol{C} = \boldsymbol{R} \cdot \boldsymbol{R} \cdot \boldsymbol{C}_{0} \cdot \boldsymbol{R}^T \cdot \boldsymbol{R}^T
\end{equation}
where, $\boldsymbol{C}$ and $\boldsymbol{C}_{0}$ denote the fourth rank elasticity tensor in the sample and crystal reference frame, respectively, and $\boldsymbol{R}$ is the rotation tensor that transforms the elasticity tensor from the crystal frame to the sample frame.


Note that although we consider the elastic anisotropy and the anisotropy factor, $a$, here, unlike crystal plasticity models, their respective evolutions are not considered in the $J_2$ plasticity model. This approximation may not be valid at large plastic strains, where the grain/crystal may rotate significantly. 

The solid solution strengthening due to alloying elements may contribute to the short range barriers to dislocation glide. In order to account for such effects, the thermal slip resistance has been formulated as \citep{sieurin2006modelling,  lindgren2017improved}:
\begin{equation}
\label{equation:s_t}
    s_t = a \left(s_t^0 + k_{LN}\ \varepsilon_b ^ {\frac{4}{3}} c_{Cr}^{\frac{2}{3}}\right)
\end{equation}
where, $a$ is the anisotropy factor described earlier (cf. Equation \ref{anisotropy_factor}), $\varepsilon_b$ accounts for the misfit strain due to solute (Cr) atoms with concentration $c_{Cr}$ and $s_t^0$ and $K_{LN}$ are associated constants. In the coupled phase field-plasticity simulations, the spatially resolved local solute concentrations are used to the solid solution strengthening model.

The evolution of the SSD density is modeled using a Kocks-Meckings type hardening equation \citep{kocks_physics_2003}, which has been modified to account for the GND density \citep{evers_non-local_2004, pai_study_2022, patra_modeling_2023}, as given below: 
\begin{equation}
\label{equation:SSD_evolution}
    \dot{\rho}_{S S D}=\frac{k_{m u l}}{b} \sqrt{\rho_{S S D}+\rho_{G N D}} \dot{\bar{\varepsilon}}^p-k_{rec} \rho_{S S D}^\alpha \dot{\bar \varepsilon}^p \\
\end{equation}
where, the first term on the RHS accounts for the dislocation evolution at forest obstacles and pre-existing junctions, while the second term accounts for the dislocation annihilation due to recovery. Here, $\rho_{GND}$ represents the GND density, and $k_{mul}$ and $k_{rec}$ are material parameters associated with the SSD density evolution.

The Nye tensor, $\boldsymbol{\Lambda}$, can be derived from the curvature of the plastic deformation gradient \citep{nye_geometrical_1953, dai1997geometrically, arsenlis_crystallographic_1999, arsenlis_evolution_2004}. We have used the same here in the rate form as \citep{patra_modeling_2023}:
\begin{equation}
    \dot{\boldsymbol{\Lambda}}=-\left(\boldsymbol \nabla \times \dot{\boldsymbol{F}}^{pT} \right)^T 
\end{equation}

The corresponding rate of evolution of GND density is given as:
\begin{equation}
    \dot{\rho}_{GND}=\frac{1}{b}||\dot{\boldsymbol{\Lambda}}||
\end{equation}
where, $||\dot{\boldsymbol{\Lambda}}||$ denotes the $L_2$ norm of $\dot{\boldsymbol{\Lambda}}$. The rate of evolution of backstress due to the GND density is given as \citep{patra_modeling_2023}:
\begin{equation}
    \label{backstress_calc}
    \boldsymbol{\dot{\chi}}=a\ k_{KH} G b \frac{\dot{\rho}_{G N D}}{2 \sqrt{\rho_{GND}}} \boldsymbol{N}^p
\end{equation}
where, $k_{KH}$ is the kinematic hardening coefficient. The incremental form of the backstress model allows development of backstress due to GNDs along the direction of plastic flow, while also capturing the history of backstress evolution upon load reversals \citep{patra_modeling_2023}. The reader is referred to \cite{pai_study_2022, patra_modeling_2023} for a detailed description of the backstress model and physical implications of the $k_{KH}$ parameter. 

In summary, we have developed a $J_2$ SGP model that can also account for the effect of crystalline anisotropy on the elastic and plastic deformation in the limit of small strains. Numerical integration of the SGP model is described in \cite{patra_modeling_2023}. The SGP model has been implemented as a material model and interfaced with the open source finite element solver, MOOSE \citep{permann_moose_2020}.

\subsubsection{Parameters for the SGP Model}
\label{model_calib}
This work primarily focuses on the rapid solidification and deformation of a model Fe-Cr alloy. In this regard, the SGP model was first calibrated to nominally predict the temperature-dependent yield stress of a representative austenitic steel \citep{nikulin_high_2010} and the representative flow stress at room temperature \citep{brown_situ_2017}. A three dimensional (3D) cube-shaped simulation domain, comprised of 512 randomly oriented cubic grains, was generated using an in-house algorithm. These 3D ensembles were meshed using hexahedral elements of $50\ \mu \mbox{m}$ size, with linear interpolation and full integration. The ensembles were deformed quasi-statically in tension at a fixed strain rate of $2 \times 10^{-4} \mbox{s}^{-1}$. Note that while the SGP constitutive model was calibrated to the experimental response using 3D simulations, the coupled PF-SGP simulations shown later have been performed in 2D, with a generalized plane strain assumption.

Flow parameters, primarily the reference strain rate, $\dot{\bar{\varepsilon}}_0^p$, activation energy for dislocation glide, $\Delta F_g$, the shape parameters, $p$ and $q$, and the threshold slip resistance, $\tau_0$, were calibrated to predict the experimentally observed yield stress as a function of temperature. Since this work involves simulating processing-induced thermo-mechanical deformation across a wide range of temperatures, temperature-dependence of the elastic constants needs to be considered as well. Temperature dependence of the anisotropic elastic constants in the range $278-1473 \mbox{K}$ was taken from \cite{magagnosc_incipient_2021, neuhaus_role_2014}. The temperature dependent thermal expansion coefficient, $\alpha^{\theta}$, was taken from \cite{pokharel_analysis_2019}. As a first order approximation, we have extrapolated these values for temperatures beyond the specified limits. FCC crystallography, with twelve ${111}<110>$ slip systems, was considered for the calculation of the anisotropy factor, $a$. 

\begin{figure}[!htbp]
    \centering
	\includegraphics[scale=0.9]{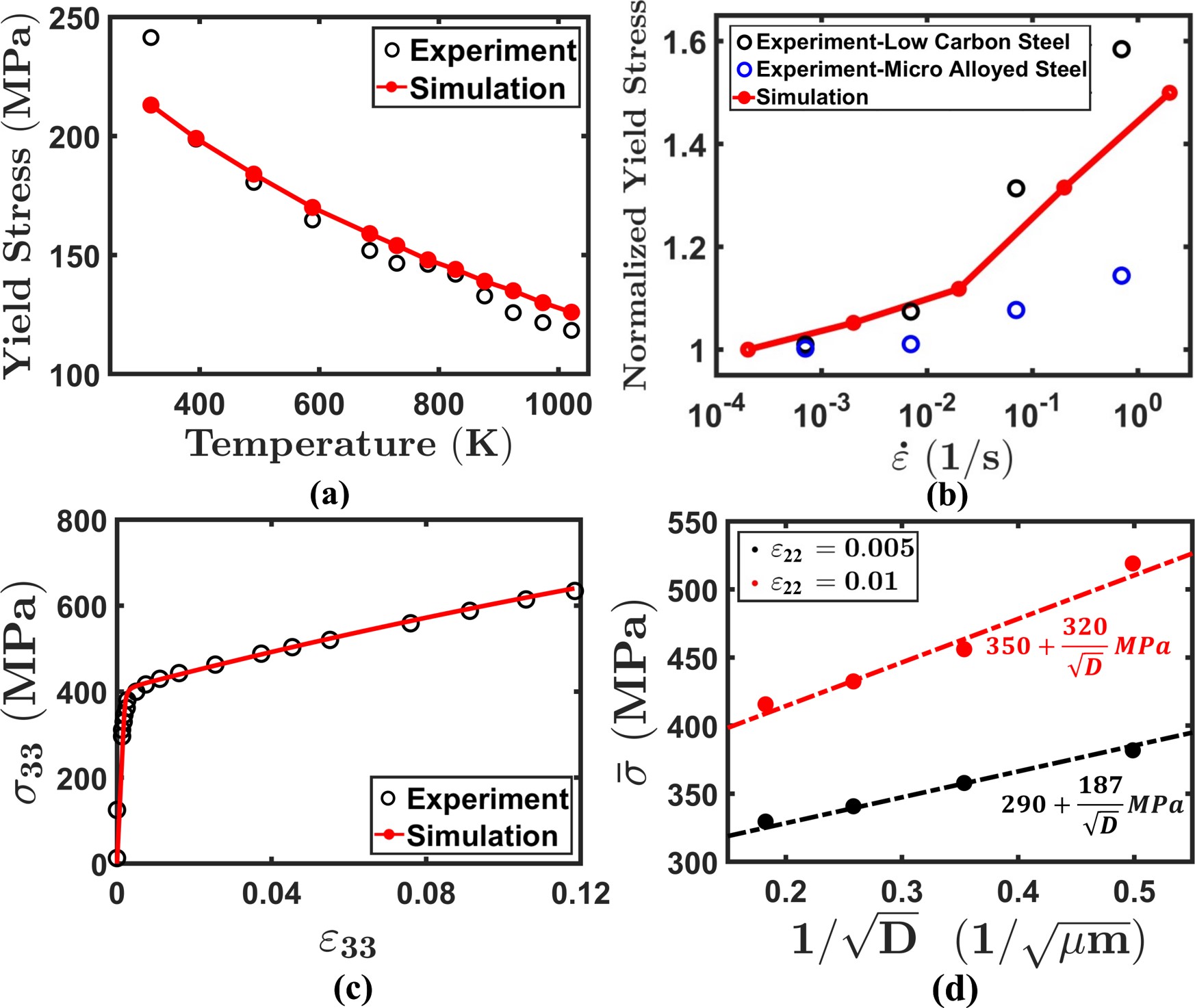}
	\caption{(a) SGP model prediction of the temperature-dependent yield stress as compared with representative experimental data from \cite{nikulin_high_2010}. (b) SGP model prediction of the normalized strain rate-dependent yield stress as compared with the representative experimental data for similar steels from \cite{paul2014tensile}. (c) Comparison of SGP predicted room temperature stress-strain response with the experimental data from \cite{brown_situ_2017}. (d) Prediction of grain size effect shown in terms of $\bar{\sigma}$ versus $1/\sqrt{D}$ for different values of applied strain. Here, $D$ represents the mean grain size in the randomly instantiated microstructures.}
    \label{fig:figure_01}
\end{figure}

Model predictions of the temperature-dependent yield stress from the standalone SGP model as compared with the experimental counterparts is shown in Figure \ref{fig:figure_01}(a). Experimental data for the temperature-dependent yield stress for SS 304L alloy were obtained from \cite{nikulin_high_2010}. Following this, the strain rate-dependent response of our SGP model was predicted by deforming the 3D ensembles at various strain rates ranging from $\dot{\varepsilon}=2\times10^{-4}\mbox{s}^{-1}$ to $\dot{\varepsilon}=2\mbox{s}^{-1}$. The $0.2\%$ offset yield strength obtained for each of the simulations, was normalized with the value obtained at the lowest strain rate, i.e., at $\dot{\varepsilon}=2\times10^{-4}\mbox{s}^{-1}$. These results, along with the experimental values for two different steels (i.e., low carbon steel: 0.035 wt. \%\ C and micro-alloyed steel: 0.062 wt. \%\ C) \citep{paul2014tensile}, have been shown in Figure \ref{fig:figure_01}(b). The model predicted strain rate sensitivity is in the same range as these representative experimental values. The hardening and substructure evolution parameters were estimated by fitting the Room Temperature (RT) stress-strain response to the experimental data for an AM SS304L deformed in tension \citep{brown_situ_2017}. Analytical estimates of the initial SSD density, $\rho_{SSD}^{0}$, and Taylor hardening coefficient for isotropic hardening, $k_{IH}$, were directly taken from \cite{brown_situ_2017}, whereas the dislocation multiplication constant, $k_{mul}$, and the dynamic recovery constant, $k_{rec}$, were estimated by fitting to the experimental RT stress-strain response. This is shown in Figure \ref{fig:figure_01}(c).

Subsequent to identifying the flow and isotropic hardening parameters, the kinematic hardening parameters were identified using a separate set of simulations in 2D polygrain ensembles. This is necessitated as the model predicted size-dependent strengthening from our first order SGP model is expected to be mesh size-dependent \citep{pai_study_2022, patra_modeling_2023}. The GND density was assumed to be negligible at the start of the calibration simulations, i.e., $\rho_{GND}^{0}=0$. The kinematic hardening coefficient, $k_{KH}$, was determined using the methodology discussed in \cite{pai_study_2022, patra_modeling_2023} to nominally predict the Hall-Petch coefficients in the range of experimental observed values. In this regard, 2D microstructure ensembles of $100\times 100\ \mu \mbox{m}$ with different mean grain sizes, $D$, were first instantiated using an open source tessellation and meshing software Neper \citep{quey2011large}. These microstructures were then meshed using 8 node square-shaped elements, with quadratic interpolation and $0.5\ \mu \mbox{m}$ mesh size. The model-predicted Hall-Petch coefficients were well within the range of experimental values reported in the literature $(\sim 320\ \mbox{MPa}\ \mu \mbox{m}^{-0.5})$ \citep{astafurov2019strain}. This has been shown in Figure \ref{fig:figure_01}(d) at different values of applied strain. Values of the SGP model parameters are given in Table \ref{table:j2_plasticity_calib}.

\begin{table}[!htbp]
   \begin{center}
     \caption{SGP model parameters for a representative Fe-Cr alloy.}
     \label{table:j2_plasticity_calib}
     \footnotesize
     \begin{tabular}{c c c}
      \hline
         Parameter & Description & Value \\ \hline
         $C_{11}$ & Temperature dependent elastic constants (GPa) & $(266.36-0.0677\theta)$ \\
         $C_{12}$ &  & $(169.6-0.0196\theta)$ \\ 
         $C_{44}$ &  & $(97.7-0.00604\theta)$ \\ 
         $G$ & Shear modulus (GPa) & $(97.7-0.00604\theta)$ \\ 
         $\alpha$ & Thermal expansion coefficient ($\mbox{K}^{-1}$) & $9.472\times10^{-6}+2.062\times10^{-8}\ \theta-8.934^{-12}\ \theta^2$ \\
         $b$ & Burgers vector magnitude (nm) & $0.258$ \\ 
         $\dot{\bar{\varepsilon}}_0^p$ & Reference shear strain rate ($\mbox{s}^{-1}$) & $1.73 \times 10^{-1}$ \\ 
         $p$, $q$ & Activation energy shape parameters & $0.35$, $1.95$ \\ 
         $\tau_{0}$ & Threshold slip resistance (MPa)& $32$ \\ 
         $\Delta F_g$ & Activation energy barrier & $0.5Gb^{3}$ \\ 
         $s_t^0$ & Thermal slip resistance (MPa) & $62.048$ \\ 
         $\varepsilon_b$ & Misfit parameter & $0.031$ \\ 
         $k_{LN}$ & Self-diffusion constant for Cr (MPa)& $18\times10^3$ \\        
         $k_{mul}$ & Dislocation multiplication constant & $0.14$ \\
         $k_{rec}$ & Dynamic recovery constant & $6$ \\
         $k_{IH}$ & Isotropic hardening coefficient due to SSDs  & $0.19$ \\
         $k_{KH}$ & Kinematic hardening coefficient due to GNDs & $0.8$ \\
         $\rho_{GND}^{0}$ & Initial GND density ($\mbox{mm}^{-2}$)& $0.0$\\
         $\rho_{SSD}^{0}$ & Initial SSD density ($\mbox{mm}^{-2}$) & $2.4\times 10^8$ \\ \hline         
     \end{tabular}
   \end{center}
\end{table}

\subsection{Coupled Phase Field and Strain Gradient Plasticity Simulations}
\label{coupled_PF_SGP}

\begin{figure}[!htbp]
    \centering
	\includegraphics[scale=0.38]{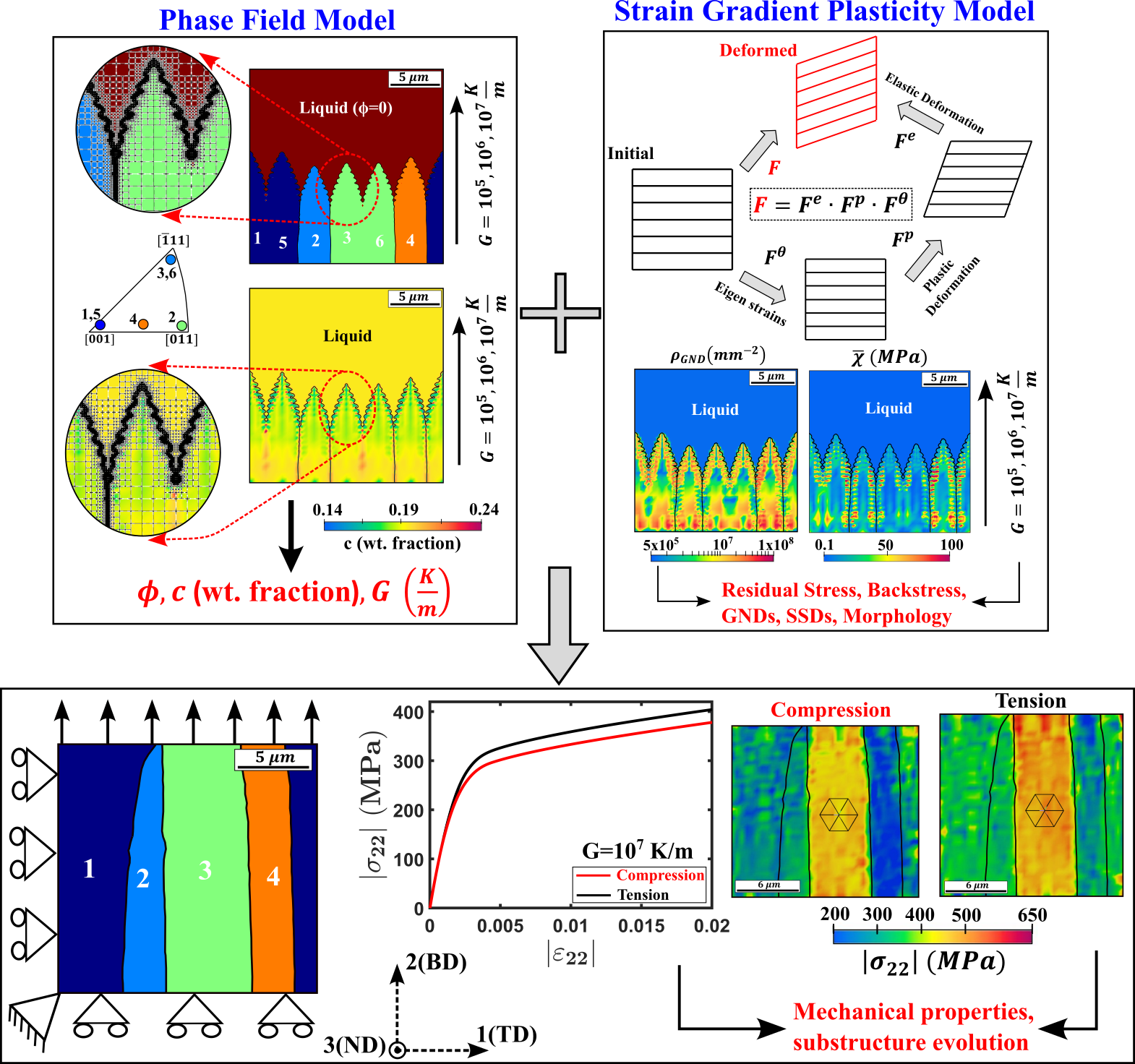}
	\caption{Schematic of the coupled phase field-strain gradient plasticity framework and its application in simulating solidification as well as post-solidification Tension-Compression asymmetry in rapidly solidified microstructures.}
	\label{fig:figure_02}
\end{figure}

Figure \ref{fig:figure_02} schematically describes our coupled PF-SGP framework. The solidification and deformation simulations have been performed in three stages: 
\begin{enumerate}
    \item The first stage involved simulation of the phase transformation due to solidification, up to the point of complete solidification, i.e., $\phi_i = 1$ in the entire simulation domain, using the coupled PF-SGP model. Consideration for SGP allows the development of dislocation substructures and residual stresses during solidification.
    \item Following complete solidification, the simulation domains are allowed to cool down to RT using only the SGP model. The PF model is turned off during this stage to save computational costs. This assumption is justified based on the limited diffusivity of Cr in solid Fe in these rapidly solidified microstructures \citep{lindroos_dislocation_2022}, such that the Cr concentration may not evolve significantly. Moreover, the primary purpose of this stage of the simulation is to allow the development of thermally-induced residual stresses (cf. \cite{pokharel_analysis_2019}).
    \item After cooling to RT, the microstructures are deformed using the SGP model, with the "residual" state of stresses, deformation gradients, SSD and GND densities, and backstresses.
\end{enumerate} 

The simulation domain of $18 \times 18\ \mu \mbox{m}$ was meshed using square-shaped finite elements, with linear interpolation and full integration. This relatively small domain size was considered primarily to mitigate computational costs (also see Section \ref{discussion}). An initial mesh size of $0.125\ \mu \mbox{m}$ was used at the start of the first time step. Following this, adaptive mesh refinement was employed to efficiently resolve the growing interface  \citep{biswas_solidification_2022}. The mesh was allowed to selectively refine around regions depicting the largest error (of all variables, conserved as well as non-conserved). Such high error values are typical for the solid-liquid interfaces, inter-dendritic regions and for regions in the vicinity of grain boundaries. This resulted in the element size varying from $0.5\ \mu \mbox{m}$ (maximum) up to $0.015\ \mu \mbox{m}$ (minimum) within the simulation domain. Since the SGP model is a non-local formulation, the element size does affect the predicted strength contribution via the kinematic hardening coefficient, $k_{kh}$ \citep{pai_study_2022,patra_modeling_2023}. As a first order approximation, we assume that the $k_{KH}$ remains constant ($k_{KH}=0.8$) over all element sizes ($0.015-0.5\ \mu \mbox{m}$) lying within the simulation domain. Following the approach of Pinomaa, Lindroos and co-authors \citep{pinomaa_significance_2020, lindroos_dislocation_2022}, six equally spaced nuclei, with an initial radius of $0.07\ \mu \mbox{m}$ and an interface width of $0.14 \mu \mbox{m}$, were initialized at the bottom edge of the simulation domain. Of the six nuclei, four have random crystallographic orientations (see orientations $1$, $2$, $3$ and $4$ in the Inverse Pole Figure (IPF) in Figure \ref{fig:figure_02}), whereas the remaining two nuclei are assumed to have an orientation identical to their neighbors (cf. nuclei $5$ and $6$ in Figure \ref{fig:figure_02}). Such an instantiation was deliberate, in order to analyze the solute behavior near Low Angle (LAGBs) or identical grain boundaries and the High Angle Grain Boundaries (HAGBs) during rapid solidification. Additionally, grains $1$ and $2$ represent the so-called intermediate orientations, having a Schmid factor $m=0.42$, whereas grain $3$ represents the hard orientation, having a $m=0.31$. The grain $4$ represents the soft orientation, having a Schmid factor of $m=0.48$, respectively.

For the seed nuclei, we have assumed an initial SSD density $(\rho_{SSD}^{0})$ of $10^2 \mbox{mm}^{-2}$ and an initial GND density $(\rho_{GND}^{0})$ of $10^2 \mbox{mm}^{-2}$. Note that these magnitudes vary from those used in our calibration simulations reported in Section \ref{model_calib}. This is because the seed nuclei are instantiated in a liquid melt pool, i.e., at $\theta \sim \theta_{m}$, when dislocation densities are expected to be low. However, the remaining parameters of the SGP model have been kept identical to those mentioned in Table \ref{table:j2_plasticity_calib}. The anisotropic strength, $\varsigma_i$ ($i\in [1,4])$, was varied to incorporate the effect of crystallographic orientation on the growth morphology. A crystal with $<001>$ axis lying closer to the thermal gradient would always have a higher $\varsigma_i$, thus exhibiting a stronger preference for growth (cf. Equation \ref{anisotropic_grain_growth}). In a similar manner, based on the crystallographic orientation allotted to the seed nuclei (see standard IPF triangle in Figure \ref{fig:figure_02}), $\varsigma_i$ calculations were performed for each of the four grains considered in this study. These values have been summarized in Table \ref{table:eps_values}. 

\begin{table}[!htbp]
   \begin{center}
     \caption{Anisotropic strength values, $\varsigma_i$, assigned to each of the grains using Equation \ref{eqn:anisotropic_interfacial_width}.}
     \label{table:eps_values}
     \footnotesize
     \begin{tabular}{c c c}
      \hline
         Grain number & Crystal Orientation (Axis) & Value \\ \hline
         $1$ & $<0\ 0\ 1>$ & $0.04$ \\
         $2$ & $<0\ 1\ 1>$ & $0.029$ \\ 
         $3$ & $<1\ 1\ 1>$ & $0.024$ \\ 
         $4$ & $<0\ 4\ 11>$ & $0.035$\\ 
         \hline         
     \end{tabular}
   \end{center}
\end{table}

The Cr concentration, $c_{Cr}$, was initialized to be $0.13$ (in wt. fraction) within the domain of the initial nuclei, while it was $0.21$  (in wt. fraction) inside the liquid. The Cr concentration was intentionally kept lower within the initial nuclei, to mimic realistic conditions, where initial nuclei formation results in solute rejection into the liquid \citep{spittle1989computer}. During the solidification step, the temperature of the top edge of the simulation cell was kept equal to the melting temperature, $\theta_m$, whereas the temperature at the bottom edge varied depending on the applied thermal gradient, $G$, and according to the following equation:
\begin{equation}
    \theta(y)=\theta_m-G(H-y)\ \forall x
\end{equation}
Here, $H = 18 \mu \mbox{m}$ is the height of the simulation domain and $y$ represents the spatial coordinate along the solidification direction ($y\in[0,18]\ \mu \mbox{m}$). The thermal gradient, $G$, varied from $10^{5}\ \mbox{K}/\mbox{m}$ to $10^{7}\ \mbox{K}/\mbox{m}$ in our simulations.

As described earlier, the phase field model operates in terms of the reduced temperature, $\theta_r$, while the deformation model operates in terms of the absolute temperature, $\theta$. In order to scale this variable, the reference temperature, $\theta_{ref}$, used in Equation \ref{eqn:reduced_temperature} to non-dimensionalize the absolute temperature, $\theta$, was kept equal to the minimum (imposed) temperature among all solidification simulations used in the present study, i.e., $\theta_{ref}=\theta_{m}-10^7\ H=1609\ \mbox{K}$. For the solidification simulations, the bottom left corner was fixed in all directions to prevent rigid body translation. Periodic boundary conditions were imposed on the left and right edges in all degrees of freedom, while the top and bottom faces were kept traction free. These non-linear partial differential equations were solved using the preconditioned Jacobian-free Newton–Krylov (PJFNK) solver in the MOOSE finite element framework \citep{permann_moose_2020}. The solver parameters and tolerances were adapted from \cite{biswas_solidification_2022}. 

Post the completion of the solidification step, i.e., $\phi_i = 1$ in the entire domain, the PF model was turned off and the mesh was regularized to a uniform mesh size of $0.5\ \mu \mbox{m}$, while ensuring smooth interpolation of the phase field variables, deformation gradient, stress tensor, SSD, GND, and the backstress tensor, from the adaptively refined mesh to the uniform mesh. Subsequent simulations were performed with the standalone SGP model for cooling the microstructures to room temperature. The microstructure solidified with the largest thermal gradient ($G=10^{7}\ \mbox{K}/\mbox{m}$), was subjected to a cooling rate of $3\times10^5\ \mbox{K}/\mbox{s}$, whereas the microstructure solidified with intermediate ($G=10^{6}\ \mbox{K}/\mbox{m}$) and low ($G=10^{5}\ \mbox{K}/\mbox{m}$) thermal gradients were cooled at $3\times10^4\ \mbox{K}/\mbox{s}$ and $3\times10^3\ \mbox{K}/\mbox{s}$ to room temperature, respectively. Such large thermal gradients, and corresponding cooling rates, were chosen to mimic realistic solidification environments which are typical in an AM process \citep{farshidianfar_effect_2016, prasad_towards_2020}. Finally, the solidified and cooled microstructures were then subjected to tensile and compressive deformations up to a nominal strain of $0.02$ to obtain their mechanical properties.

\section{Model Predictions}
In the present study, solidification simulations have been performed for three different thermal gradients, namely, $10^5\ \mbox{K}/\mbox{m}$, $10^6\ \mbox{K}/\mbox{m}$, and $10^7\ \mbox{K}/\mbox{m}$. We first present model predictions of microstructure evolution during solidification, followed by the mechanical properties.

\subsection{Evolution of the conserved parameter, $c_{Cr}$}
\label{conc_distribution}

\begin{figure}[!htbp]
    \centering
	\includegraphics[scale=0.85]{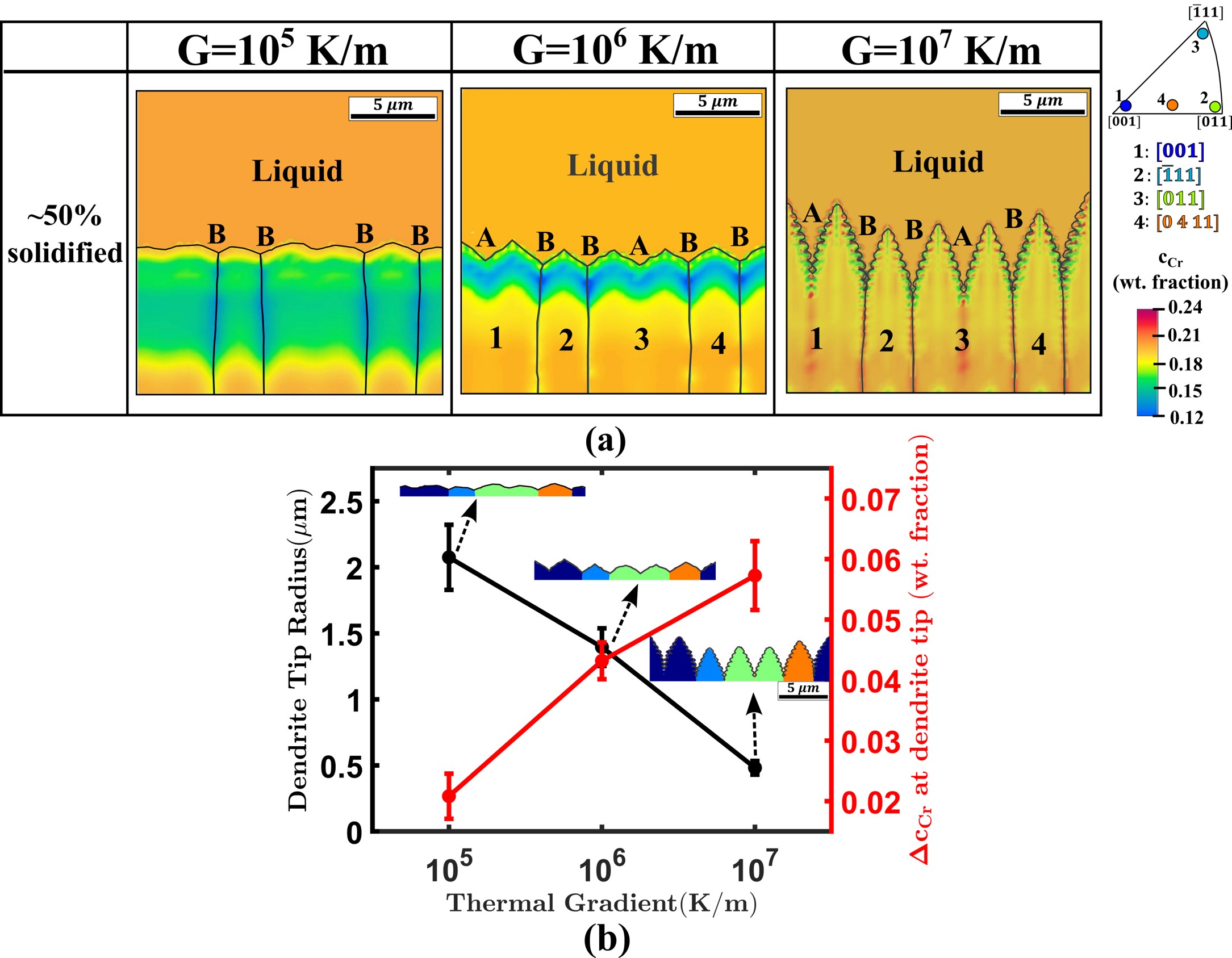}
	\caption{Cr concentration, $c_{Cr}$, in the $\sim 50\%$ solidified microstructure for three different thermal gradients, $G$. The crystallographic orientations of the grains have been marked in the standard inverse pole figure (IPF) key. The boundaries at the grain interfaces have been added during post-processing. The markers ‘A’ and ‘B’ refer to identical (same orientation between grains) and high angle grain boundaries, respectively. (b) Effect of the imposed thermal gradient, $G$, on the radius and Cr solute segregation at the dendrite tip.}
	\label{fig:figure_03}
\end{figure}

Figure \ref{fig:figure_03} shows the distribution of Cr solute, $c_{Cr}$ (in wt. fraction), after $\sim 50\%$ solidification for three different thermal gradients. The crystallographic orientation of the individual grains has been highlighted in the corresponding standard inverse pole figure (IPF) key. Two important observations emerge from these contours. Firstly, a transition from cellular to dendritic grain morphology can be seen with increasing values of $G$ from $10^5\ \mbox{K}/\mbox{m}$ to $10^7\ \mbox{K}/\mbox{m}$. Such grain morphologies are typical in welded structures, where a sharp thermal gradient, $G$, leads to a planar-cellular-dendritic transition on moving from the base metal towards the weld region \citep{zhu2021planar, fisher2023fundamentals}. Further, the solidification front is non-uniform, i.e., grains oriented favorably tend to grow faster than their neighbors. The anisotropy strength, $\varsigma_i$, plays a key role here. Grains with $<001>$ axis oriented closer to the thermal gradient direction have a higher value of anisotropy strength, thus giving them a competitive advantage over their neighbors \citep{liu_investigation_2018}. However, it is important to note that this heterogeneity in grain morphology is observed only for large thermal gradients. Lower thermal gradients provide a diminished driving force for grain growth, thus reducing the competitive advantage of favorably oriented grains and resulting in a cellular front. 

Identical grain boundaries (same orientations between the grains) are marked as ‘A’, while the HAGBs are marked as ‘B’ in Figure \ref{fig:figure_03}. It is observed that the Cr segregation at ‘A’ type boundaries is quite prominent for the largest $G$, i.e., $10^7\ \mbox{K}/\mbox{m}$. However, such a prominent Cr segregation does not develop at the ‘A’ type boundaries for $G = 10^5\ \mbox{K}/\mbox{m}$ and $G = 10^6\ \mbox{K}/\mbox{m}$. This is primarily because the lower thermal gradients associated with lower cooling rates provide sufficient time and thermal activation for the Cr solute for diffusion, thus leading to segregation at only ‘B’ type grain boundaries \citep{mun2012effects, li2022effect}. For the microstructures solidifying at thermal gradients of $G=10^5\ \mbox{K}/\mbox{m}$ and $G=10^6\ \mbox{K}/\mbox{m}$, the temperatures in these microstructures remains close to the melting point ($\theta_m=1789\ \mbox{K}$) even after complete solidification. Hence, no significant solute segregation can be seen at ‘A’ as well ‘B’ type boundaries in Figure \ref{fig:figure_03}. Another observation from the contours shown in Figure \ref{fig:figure_03}(a) is the variation in the Cr solute distribution within the solidified grains (along the direction of thermal gradient), specifically at low and intermediate thermal gradients. Such a behavior can arise by neglecting the effects of interface velocity on the partition coefficient \citep{pinomaa_significance_2020}, thus altering the evolution of (spatial as well temporal) Cr solute distribution (see Equation \ref{eqn:c_evol}). 


In order to further explore the effect of cooling rates on the microstructure evolution, the dendrite tip radius was estimated for the $\sim 50\%$ solidified microstructures. Figure \ref{fig:figure_03}(b) shows a decrease in the dendrite tip radius with increasing cooling rate. The observed scatter is primarily due to the anisotropy associated with the randomly assigned crystallographic orientation of the growing nuclei. The dendrite tip radius provides an alternative indication of the wavelength of instability at the solid-liquid interface \citep{fisher2023fundamentals}. Moreover, large thermal gradients, associated with high cooling rates result in a significant undercooling at the solid-liquid interface. This is due to the imposed temperature boundary conditions that alter the local temperature as well as composition equilibrium from the calculated solidus temperature, and are primarily responsible for the complex non-equilibrium microstructures observed in rapidly solidified microstructures \citep{zhu2021planar}.

Further, the Cr diffusivity in the solid phase is significantly lower as compared to the liquid phase. This, in conjunction with the higher undercooling and enhanced boundary layer thickness, leads to a higher $\Delta Cr$ at the solid-liquid interface for simulations with large thermal gradients (and hence higher cooling rates). This phenomenon has been quantified in Figure \ref{fig:figure_03}(b), by extracting the Cr concentration from the line profiles at the solid-liquid interface from the $\sim 50\%$ solidified microstructure. 

The results presented in this section highlight the effect of thermal gradients on solute (Cr) partitioning to the grain boundaries. Presumably, these may be expected to affect the deformation behavior and mechanical properties as well. In order to account for such effects, the thermal slip resistance in our SGP model accounts for solid solution strengthening due to the local Cr concentration (cf. Equation \ref{equation:s_t}). Based on the local Cr concentration, the local thermal slip resistance is also allowed to vary during solidification.

\subsection{Evolution of GND Density and Backstress}
\label{GND_dist}

\begin{figure}[!htbp]
    \centering
	\includegraphics[scale=0.4]{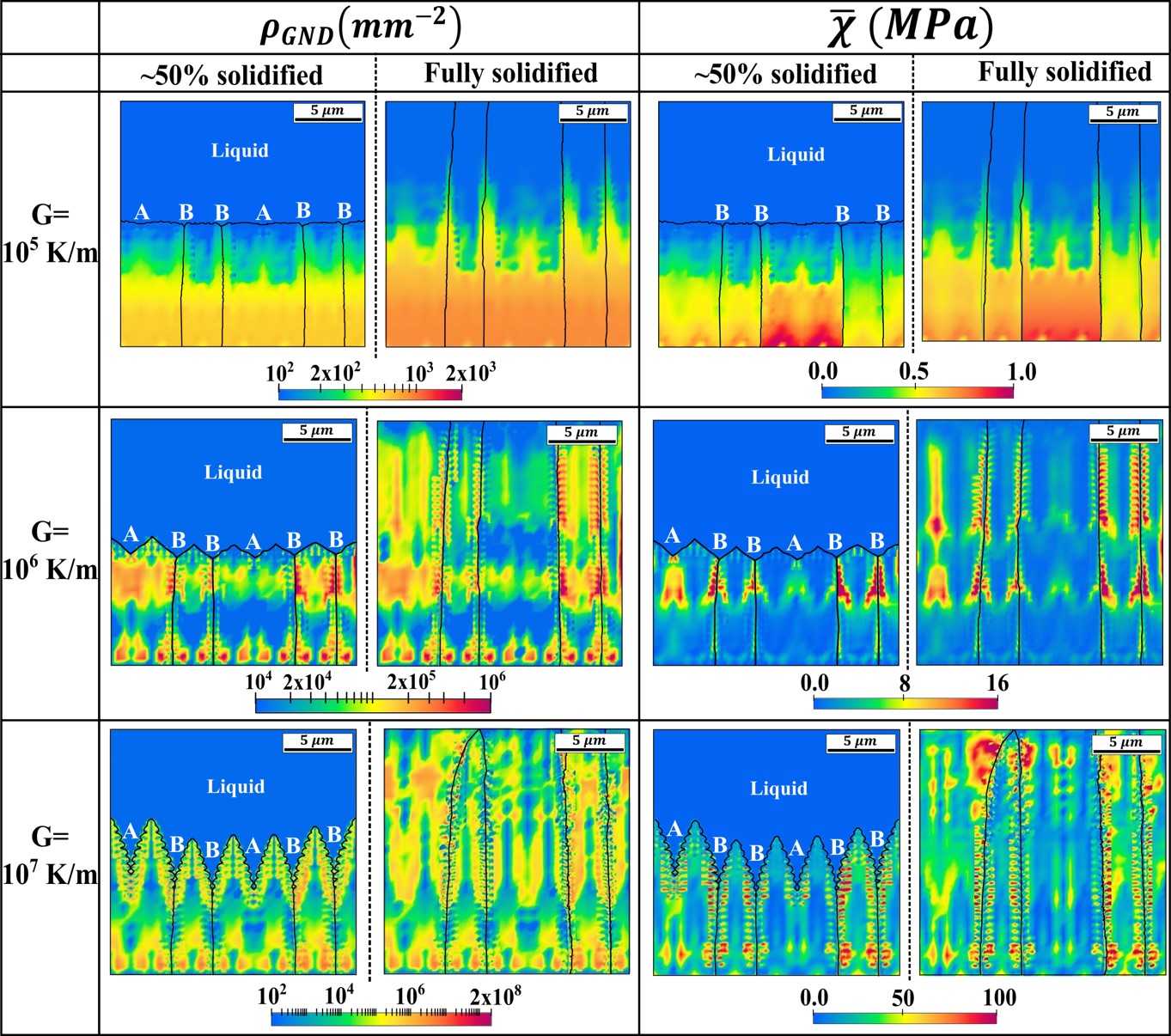}
	\caption{Contours of the GND density, $\rho_{GND}$, and effective backstress, $\bar{\chi}$, in the $\sim 50\%$ and the fully solidified microstructures for three different thermal gradients.}
	\label{fig:figure_04}
\end{figure}

\begin{figure}[!htbp]
    \centering
	\includegraphics[scale=0.55]{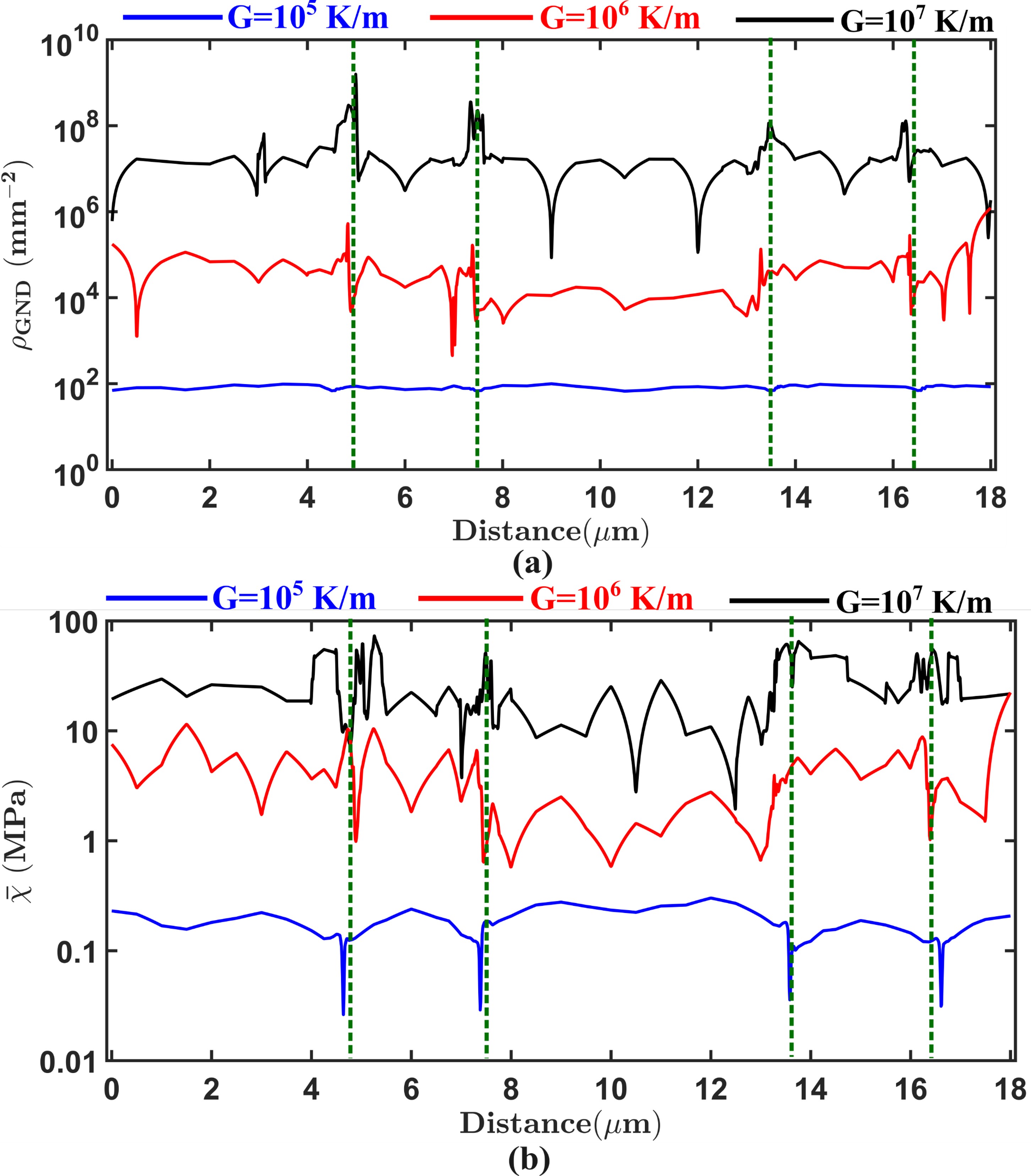}
	\caption{Line profiles of (a) GND density, $\rho_{GND}$, and (b) effective backstress, $\bar{\chi}$, in the fully solidified microstructures at $H/2$ for simulations with three different thermal gradients.}
	\label{fig:figure_05}
\end{figure}

\begin{figure}[!htbp]
    \centering
	\includegraphics[scale=0.55]{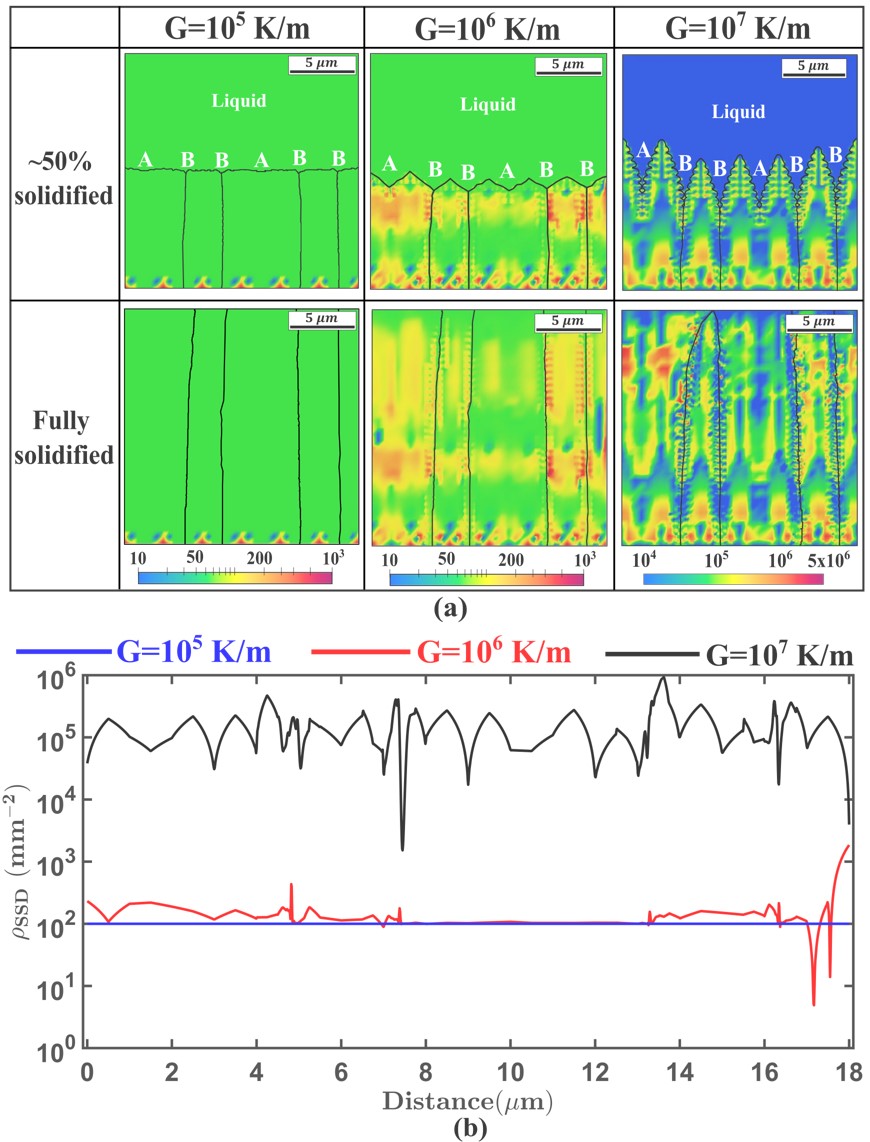}
	\caption{(a) Contours of the SSD density, $\rho_{SSD}$, in the $\sim 50\%$ and the fully solidified microstructures for three different thermal gradients. (b) Line profiles of the SSD density in the fully solidified microstructures at $H/2$ for simulations with three different thermal gradients.}
    \label{fig:figure_06}
\end{figure}

Figure \ref{fig:figure_04} shows the evolution of the GND density, $\rho_{GND}$, and the effective backstress, $\bar{\chi}$, in the $\sim 50\%$ and the fully solidified microstructures under three different (imposed) thermal gradients. Corresponding line profiles for $\rho_{GND}$ and $\bar{\chi}$ obtained at $H/2$ for the fully solidified microstructures are shown in Figures \ref{fig:figure_05}(a) and \ref{fig:figure_05}(b), respectively. In addition, the contour plots and the corresponding line profiles at $H/2$ for the SSD density, $\rho_{SSD}$, have been shown in Figure \ref{fig:figure_06}(a) and \ref{fig:figure_06}(b). Localizations of $\rho_{GND}$, and the corresponding $\bar{\chi}$ are evident along the primary as well as secondary dendrite arms for the microstructure solidified at $G=10^{7}\ \mbox{K}/\mbox{m}$. Insufficient supply of liquid metal to fill the empty gaps between the dendritic arms results in the development of significant strain gradients within these regions. The Cr solute segregation, especially at larger imposed thermal gradients, in the dendrite arms adds up to the local microstructural heterogeneity as well (cf. Figure \ref{fig:figure_03}). This leads to the development of geometrically necessary dislocations in these regions.

An important observation from Figure \ref{fig:figure_04} and \ref{fig:figure_05}(b) is the development of backstress along the ‘A’ type boundaries. Especially, the microstructure solidified under the largest thermal gradient shows a significant backstress accumulation (Figure \ref{fig:figure_04}) and solute segregation (cf. Figure \ref{fig:figure_03}) along the ‘A’ type boundaries. At lower thermal gradients, the microstructure tends to homogenize, thus diffusing out the $\rho_{GND}$ (Figure \ref{fig:figure_05}(a)) and backstress (Figure \ref{fig:figure_05}(b)) accumulations. This indicates that the development of local GNDs and backstress is also dependent on the underlying solute segregation. Generally speaking, the grain boundary energy associated with identical grain boundaries is significantly lower than their HAGB counterparts \citep{humphreys2012recrystallization}. Hence, the heterogeneities developed across ‘A’ type boundaries depict a much lower magnitude than the ones developed across HAGBs (see Figure \ref{fig:figure_04}). With decrease in $G$, no significant GND density and backstress accumulations were observed along the ‘A’ type boundaries for the intermediate and low thermal gradients (cf. Figure \ref{fig:figure_05}). 

It can be seen from Figure \ref{fig:figure_05} that the GND density increases by $\sim 1-2$ orders of magnitude and the effective backstress increases by about an order of magnitude per order of magnitude increase in $G$. While the value of backstress is still insignificant ($< 6\ \mbox{MPa}$) for thermal gradients of $10^6\ \mbox{K}/\mbox{m}$ and lower, the values increase to $\sim 30\ \mbox{MPa}$ for a thermal gradient of $10^7\ \mbox{K}/\mbox{m}$, which is typical of AM processes. \cite{bertsch2020origin} measured the dislocation density in TEM lamellae of AM SS316L, where the cooling rate were derived from the corresponding SDAS spacing. Similar to the present work, they have observed a net rise in the total dislocation density by $\sim 2$ orders of magnitude with increase in the cooling rate from $3\times10^3\ \mbox{K}/\mbox{s}$ to $7\times10^4\ \mbox{K}/\mbox{s}$. Additionally, similar magnitudes of total dislocation density have also been reported by \cite{hu2023dislocation} at the end of the cooling stage using their multi-scale modeling framework of laser-based AM process. \cite{yoo2018identifying} calculated High Resolution- Electron Back Scattered Diffraction (HR-EBSD) based GND densities in an AM Inconel 718, and reported GND localizations similar to those shown in Figure \ref{fig:figure_04} in the inter-dendritic regions. \cite{godec2020quantitative} reported an average GND density of $1.34\times10^7\ \mbox{mm}^{-2}$ in their AM SS316L from the EBSD based Kernel Average Misorientation (KAM) data. Similar magnitudes of GND densities can also be seen in Figure \ref{fig:figure_04} for the specimen solidifying with the highest $G$. \cite{small2021role} reported a rise in EBSD based average GND densities from $2.2\times10^{7}\ \mbox{mm}^{-2}$ to $1.2\times10^8\ \mbox{mm}^{-2}$ on moving from Binder Jet 3D Printing (BJ3P), with low thermal gradient, to Direct Metal Laser Sintered (DMLS), with higher thermal gradient for AM Inconel 625. Such an effect of the imposed thermal gradient can be clearly observed in Figure \ref{fig:figure_04} and \ref{fig:figure_05}(a). The above observations provide validation of the model predicted dislocation densities near the inter-dendritic and grain boundaries.

The $\rho_{SSD}$ contours and corresponding line profiles shown in Figure \ref{fig:figure_06}(a) and \ref{fig:figure_06}(b) do not show any notable variation for the lower and intermediate $G$, with their average magnitude being close to that of the initial SSD density, $10^2\ \mbox{mm}^{-2}$. Variations are somewhat visible at ‘A’ as well as ‘B’ type boundaries for the microstructure solidified under $G=10^6\ \mbox{K}/\mbox{m}$, however the subtended magnitudes are not that significant. On the other hand, the microstructure solidified under $G=10^7\ \mbox{K}/\mbox{m}$ does indicate a notable rise in the $\rho_{SSD}$, although the average magnitude is still two orders lower than its $\rho_{GND}$ counterpart. It is important to note here that the exponential rise in $\rho_{GND}$ with increasing $G$ also influences the $\rho_{SSD}$ evolution through Equation \ref{equation:SSD_evolution}. The sudden in rise in $\rho_{SSD}$ for microstructure solidified under $G=10^7\ \mbox{K}/\mbox{m}$, and similarity in contour plots shown in Figure \ref{fig:figure_04} and \ref{fig:figure_06}(a) do conform the above conclusion. This is interesting, considering that although our SGP model is rate-sensitive (cf. Figure \ref{fig:figure_01}(b)), the SSD density, which is a function of the plastic strain, does not vary significantly for the different thermal gradients. Given that the GND density is much higher than the SSD density for the highest $G$, the dislocation structures could be expected to be primarily of GND type. These results also indicate that the thermal gradient induced GND density and backstress could not be predicted by \textit{local} plasticity models that do not account for the effect of strain gradients. \textit{Non-local} strain gradient plasticity formulations are required to simulate the heterogeneous microstructure and residual stress developments during rapid solidification. 

\subsection{Uniaxial Deformation of Solidified Microstructures}
\label{uni_def}

The solidified microstructures were first cooled to room temperature and then loaded in uniaxial tension and compression along the build (BD) and transverse (TD) directions at a quasi-static strain rate of $2\times10^{-4}\ \mbox{s}^{-1}$ up to $0.02$ nominal strain. As explained earlier, the microstructure solidified with the largest thermal gradient ($G=10^7\ \mbox{K}/\mbox{m}$) was cooled to RT with a cooling rate of $3\times10^5\ \mbox{K}/\mbox{s}$, whereas its lower ($G=10^5\ \mbox{K}/\mbox{m}$) and intermediate ($G=10^6\ \mbox{K}/\mbox{m}$) counterparts were subjected to a cooling rate of $G=3\times10^3\ \mbox{K}/\mbox{s}$ and $G=3\times10^4\ \mbox{K}/\mbox{s}$. As shown in Figure \ref{fig:figure_02}, the bottom and left faces were subjected to roller boundary conditions, whereas the bottom left corner was fixed in all dimensions to avoid rigid body motion. The remaining surfaces were kept traction free. As discussed previously in Section \ref{coupled_PF_SGP}, the PF model was turned off and only the SGP model was employed for these deformation simulations.

\begin{figure}[!htbp]
    \centering
	\includegraphics[scale=0.45]{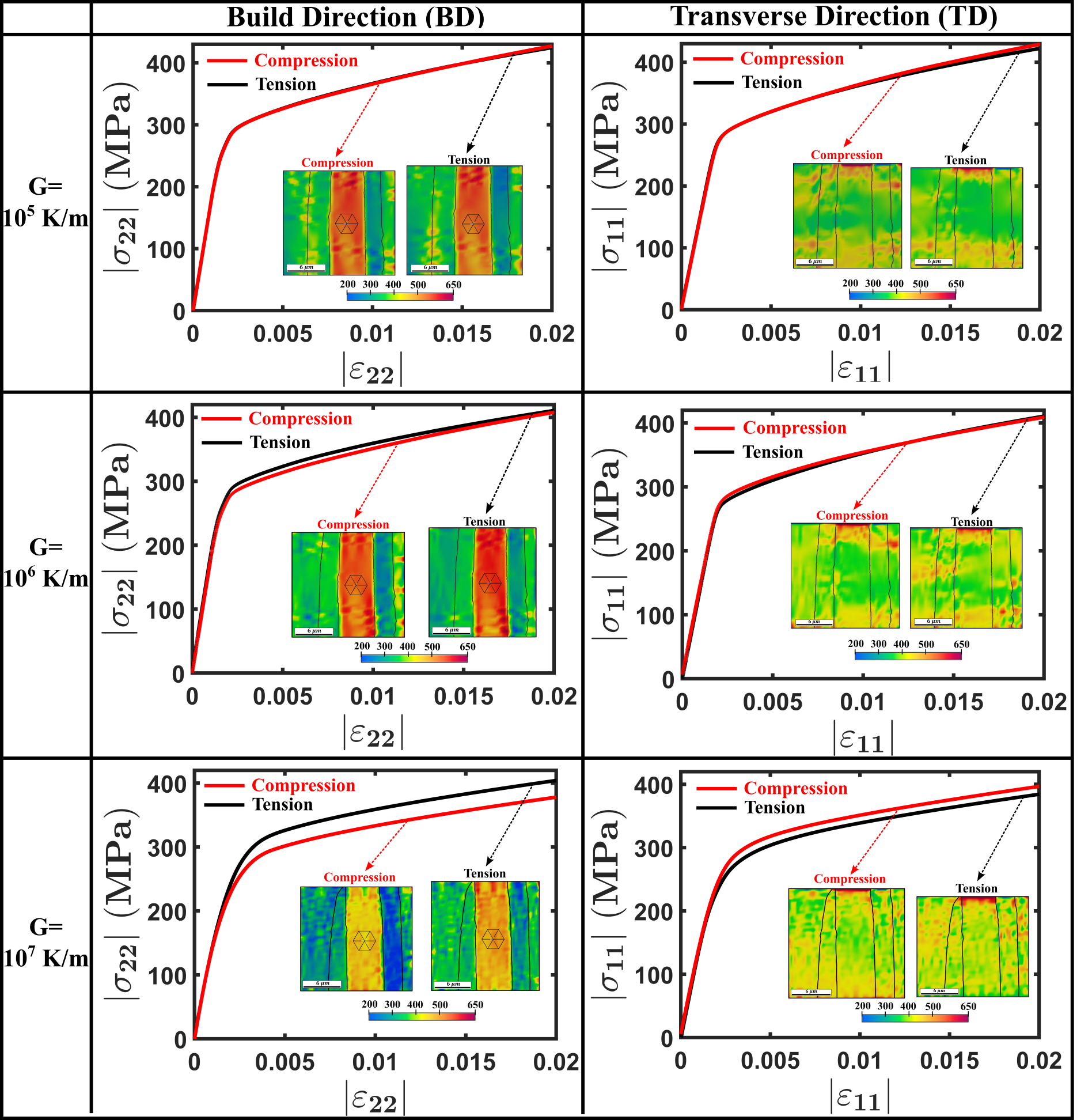}
	\caption{Stress-strain responses of the microstructures solidified andcooled under three different thermal gradients, loaded in uniaxial tension and compression along the build direction (BD) and the transverse direction (TD). The $|\sigma_{22}|$ (for BD loading) and $|\sigma_{11}|$ (for TD loading) deformation contours at a nominal strain of $0.02$ have been added as an inset for each of the thermal gradients, respectively.}
	\label{fig:figure_07}
\end{figure}

Figure \ref{fig:figure_07} shows the predicted mechanical response when loaded in both tension and compression, and separately along the BD and TD, for the microstructures solidified (and further cooled) with different thermal gradients. Generally speaking, the flow stress is higher during compression loading as compared to tension loading for the microstructures loaded along the TD and vice-versa for those loaded along the BD. Moreover, the observed difference in flow stress between tension and compression increases with increase in the (imposed) thermal gradient $G$. For $G = 10^5\ \mbox{K}/\mbox{m}$, this effect is almost negligible. With a thermal gradient of $10^6\ \mbox{K}/\mbox{m}$, this effect increased somewhat, but is still very small. However, under large thermal gradients $(G=10^7\ \mbox{K}/\mbox{m})$, the anisotropy in predicted mechanical response is clearly visible. This behavior could be attributed to the significant $\rho_{GND}$ and backstress localizations developing in the microstructure, as seen in Figure \ref{fig:figure_04} and is discussed later. Further, no appreciable difference is observed in the work-hardening regime for the different microstructures. This could possibly be due to the observation that $\rho_{SSD}$ spatial distribution was nearly homogeneous, and did not display any significant localizations for the different thermal gradients (cf. Figure \ref{fig:figure_06}(a)) used in this study. The contours of $|\sigma_{22}|$ (for BD loading) and $|\sigma_{11}|$ (for TD loading), shown as inset in Figure \ref{fig:figure_07}, show an appreciable difference in the magnitude as well as distribution between the tensile and compressive deformations for the highest $G$. This is observed for the BD as well as TD loaded microstructures. While this effect is still present at lower $G$, the difference in the stress magnitudes between the two loadings are relatively lower. 

In addition, the hard grain ($m=0.31$, grain $3$ in Figure \ref{fig:figure_02}) displays significantly large stress magnitudes as compared to the intermediate grains ($m=0.42$, orientation $1$ and $2$ in Figure \ref{fig:figure_02}), which in turn, accommodate higher stresses when compared to the softer grain ($m=0.49$, orientation $4$ in Figure \ref{fig:figure_02}). Such an effect is much more prominent when loaded along the BD than along the TD. Similar correlations between the accommodated residual stresses and (grain average) orientations have been noted by employing a thermo-mechanical crystal plasticity model during LPBF-based solidification of a fully austenitic SS316L by \citep{kuna2023framework}. These results further indicate that the proposed SGP model has the ability to capture orientation-based anisotropy in directionally solidified microstructures at a computational cost significantly lower than its CPFE counterpart. Additionally, the Cauchy stress contours for the different solidified and cooled microstructures are presented in Appendix \ref{sec:appendix_B} to verify the self-equilibration of the residual stresses.

\begin{figure}[!htbp]
    \centering
	\includegraphics[scale=0.6]{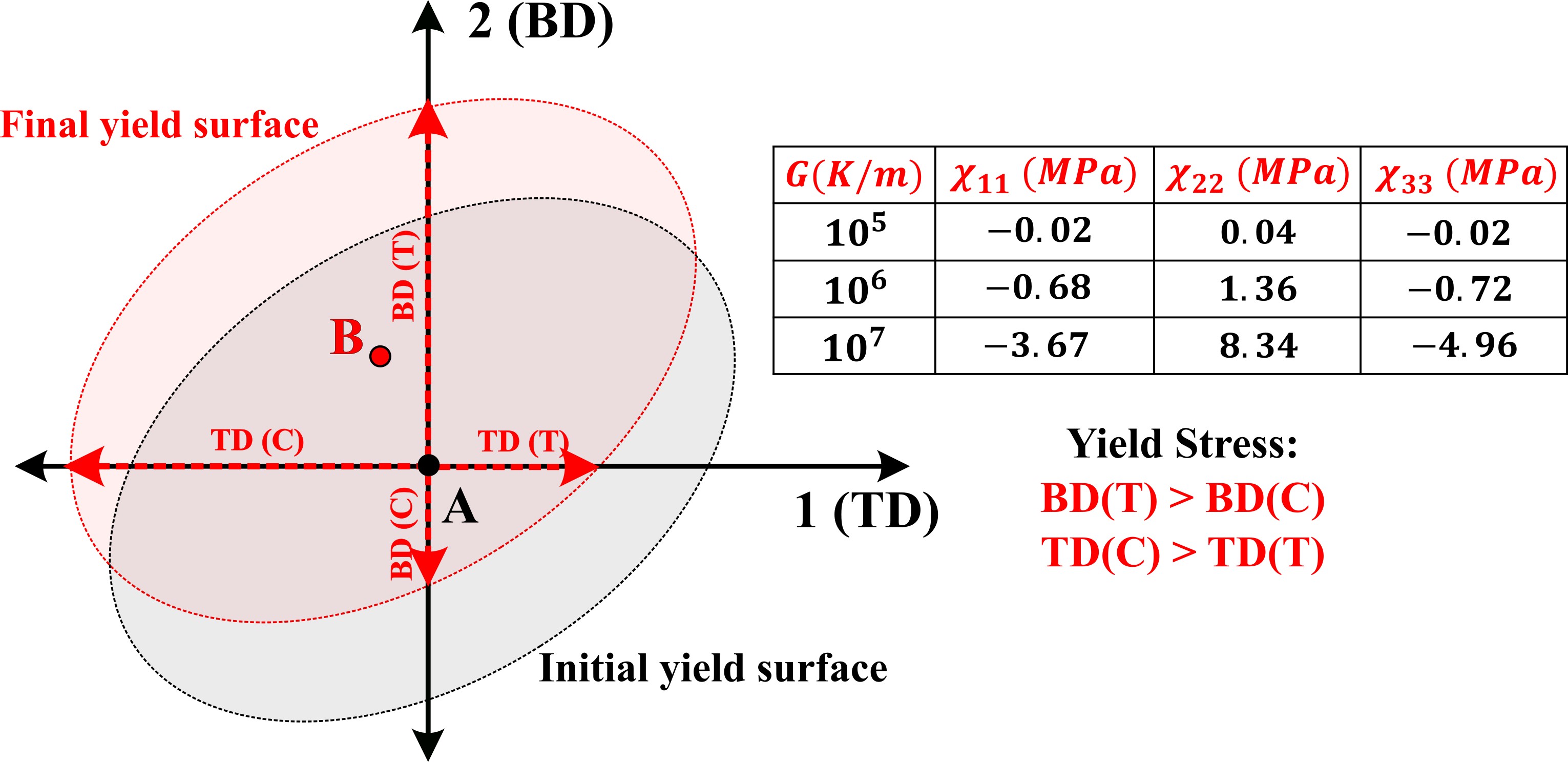}
	\caption{Schematic showing the effect of the backstress tensor, $\boldsymbol{\chi}$, on the translation of the yield surface due to dislocation substructure development during solidification. The table on the right gives the values of the individual components of the backstress tensor, $\chi_{11}$, $\chi_{22}$ and $\chi_{33}$, at the end of solidification and cooling for the microstructures with three different thermal gradients.}
	\label{fig:figure_08}
\end{figure}

To further understand the role played by printing-induced backstresses on the anisotropy in mechanical response, the average values of individual backstress components, i.e., $\chi_{11}$, $\chi_{22}$ and $\chi_{33}$, were analyzed at the end of the cooling regime. This is described schematically in terms of the translation of the von-Mises yield surface in Figure \ref{fig:figure_08}. The tabulated (average) backstress magnitudes show a clear increase with increase in thermal gradients, irrespective of the component under consideration. Using a two dimensional approximation of 3D yield surface, we now try to explain the effect of these (directional) backstress components on the subsequent yield surface. Assuming an initially negligible backstress (which is conventionally assumed to be the center of the yield surface), the initial yield surface (schematically shown using the dotted black line in Figure \ref{fig:figure_08}) is centered at the origin, A. Dislocation substructure and backstress evolution during solidification leads to the translation of the center of the yield surface to the point B, which also varies as a function of the thermal gradient. The negative values of $\chi_{11}$ leads to a leftward translation of the yield surface along direction 1 (TD) (see Figure \ref{fig:figure_01}), along with a upward translation along direction 2 (BD) due to a positive $\chi_{22}$. The post-solidification yield surface is schematically shown using the dotted red line in Figure \ref{fig:figure_08}. On subjecting to uniaxial deformation, the resulting yield stress is thus higher in compression as compared to tension for TD loading, and vice-versa for BD loading. This explains the mechanistic origin of TC asymmetry in rapidly solidified Fe-Cr alloys due to the GND-induced directional backstress. 

Given that a generalized plane strain assumption was used in our mechanics model, $\chi_{33}$ is also predicted in our essentially 2D simulations (which is similar in magnitude to $\chi_{11}$). However, the qualitative arguments regarding the translation of the yield surface due to the printing-induced directional backstress would not change under the assumption of different boundary conditions, such as plane stress or full 3D.

\begin{figure}[!htbp]
    \centering
	\includegraphics[scale=0.45]{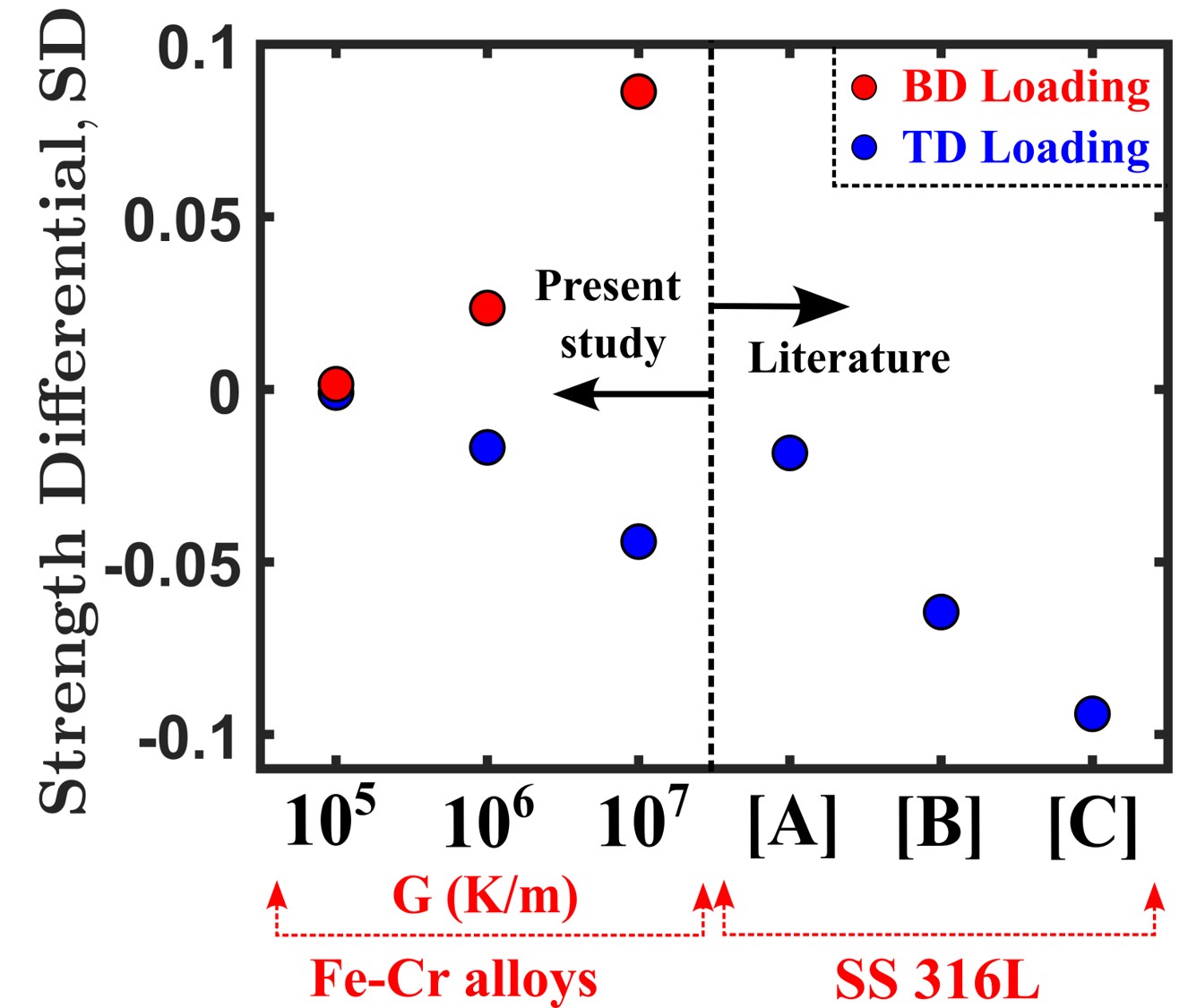}
	\caption{Strength Differential (SD) values for the three thermal gradients under BD and TD loading, in comparison with the available experimental data. The markers [A], [B] and [C] refer to the literature values from \cite{wang_anisotropic_2023}, \cite{jeon2019micro}, and \cite{chen_microscale_2019}, respectively.}
	\label{fig:figure_09}
\end{figure}

In order to quantify the predicted TC asymmetry, we have adopted the metric of Strength Differential, SD, as given by \citep{bassani2011non, patra2014constitutive}:
\begin{equation}
    S D=\frac{|\sigma_y^t| - |\sigma_y^c|}{\left(|\sigma_y^t| + |\sigma_y^c|\right) / 2}
\end{equation}
where, $\sigma_{y}^{t}$ and $\sigma_{y}^{c}$ denote the yield strength in tension and compression, respectively. Figure \ref{fig:figure_09} shows a direct comparison of the SD obtained under BD and TD loading for the various thermal gradients simulated in this study. The SD values indicate that the TC asymmetry gets enhanced with increase in the thermal gradients. Available SD data from the literature for similar Fe-Cr alloys (under TD loading) have also been plotted in Figure \ref{fig:figure_09}. It can be seen that the predicted SD values, under TD loading, are qualitatively comparable to the experimental SD values reported in literature, thus validating our model predictions. It should be noted that the present study used only six initial seeds with (four) random crystallographic orientations (see Figure \ref{fig:figure_02}), for the (directional) solidification simulations. The number of seeds used, their orientations, the imposed thermal gradient and the volume of the ensemble may also effect the ensuing mechanical properties, and hence the SD magnitudes. Further, it should also be noted that our model was not calibrated to any specific material data and hence only qualitative trends should be inferred from our predictions.

Two important observations can be further derived over here. Firstly, the microstructures solidified (and further cooled) under the intermediate and largest thermal gradients have predicted a non-negligible TC asymmetry, with the simulated SD values being qualitatively similar to those reported in the literature \citep{jeon2019micro, chen_microscale_2019, wang_anisotropic_2023}. For the TD loading of the microstructure with $G=10^7\ \mbox{K}/\mbox{m}$, the yield stress values in tension were $293.1\ \mbox{MPa}$ and compression were $305.1\ \mbox{MPa}$, respectively. This resulted in a SD of $-0.0401$. Note here that the difference between the yield stresses in compression and tension ($\sim 12.0\ \mbox{MPa}$) roughly correspond to twice of the mean effective (average) backstress accumulated in the microstructure ($\sim 7.4\ \mbox{MPa}$). \cite{chen_microscale_2019} also reported similar observations from their deformation studies on AM SS 316L. An identical analysis for the microstructure with $G=10^6\ \mbox{K}/{m}$ reveals that the difference between the yield stress in compression and tension ($\sim 4.0\ \mbox{MPa}$) follows a similar relationship with the effective (average) backstress ($\sim 1.7\ \mbox{MPa}$). Moreover, with increasing thermal gradients, the SD values display an exponential increase, as seen in Figure \ref{fig:figure_09} for BD as well as TD loading scenarios. This could be attributed to the order of magnitude increase in backstress values (due to the printing-induced GND density) as described in Figure \ref{fig:figure_08}. Presumably, local solute segregation may also play a role in the TC asymmetry of rapidly solidified microstructures. Individual contributions of these competing mechanisms are delineated in Section \ref{discussion}.


\section{Discussion}
\label{discussion}
It has been previously reported that the microscale intragranular residual stresses, an outcome of printing induced dislocation substructures, are primarily responsible for the TC asymmetry observed in rapidly solidified metals \citep{chen_microscale_2019, zhang_modeling_2022}. The present work developed a coupled PF – SGP framework to simulate these microscale residual stresses and their effect on the anisotropic mechanical properties, specifically the TC asymmetry. The prior studies  \citep{chen_microscale_2019, zhang_modeling_2022}, assumed a constant initial value of backstress during the deformation simulations to predict this TC asymmetry. In our work, the backstress, due to the GND densities, were allowed to develop \textit{in situ} during the solidification. While Pinomaa, Lindroos and co-authors \citep{pinomaa_significance_2020, pinomaa_phase_2020, lindroos_dislocation_2022} have previously used coupled phase field-crystal plasticity models to predict the dislocation structures and also predicted the mechanical properties of rapidly solidified microstructures, they did not explicitly model the \textit{directional} backstress due to these dislocation structures and did not attempt to predict the anisotropic mechanical properties of rapidly solidified microstructures. Consideration for the said effects in our coupled PF-SGP modeling framework to predict the TC asymmetry of rapidly solidified microstructures represents an advancement over these prior modeling studies.

\begin{figure}[!htbp]
    \centering
	\includegraphics[scale=0.5]{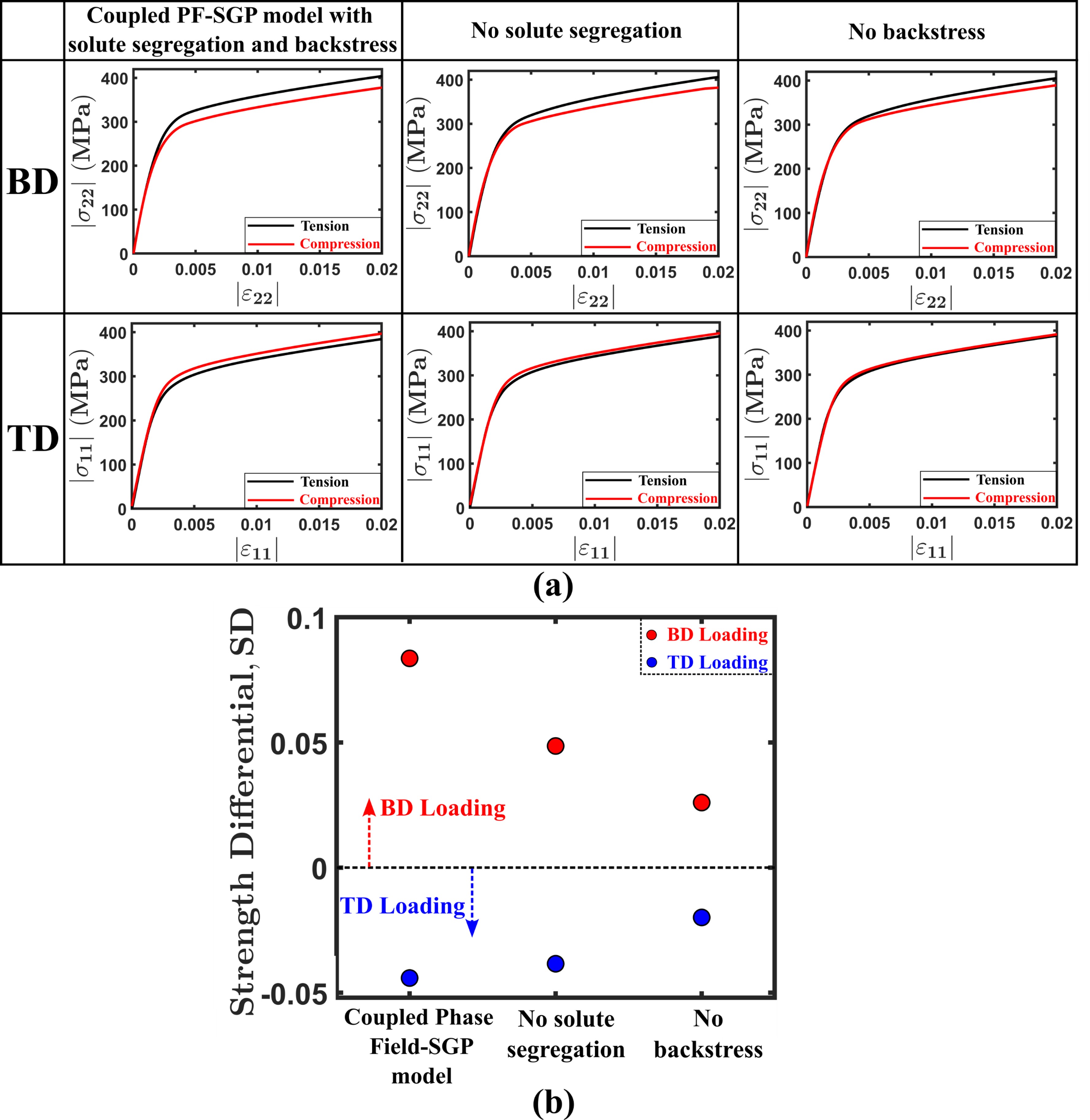}
	\caption{(a) Predicted stress-strain responses along the BD and TD, and (b) Strength Differential, SD, values for microstructures solidified under $G = 10^7\ \mbox{K}/\mbox{m}$ and then cooled with a cooling rate of $3\times10^5\ \mbox{K}/\mbox{s}$ using the fully coupled PF-SGP model (with solute segregation and directional backstress), and those without consideration for solute segregation, and backstress, respectively.}
	\label{fig:figure_10}
\end{figure}

We further delineate the role played by the two potentially competing mechanisms on the TC asymmetry: (a) the thermal distortion induced GNDs (and directional backstress), and (b) local Cr solute segregation at the solid-liquid interface, which necessitates the presence of GNDs to accommodate the ensuing lattice incompatibilities. As discussed in \citep{yoo2018identifying, bertsch2020origin}, these solute segregation-induced GNDs then interact with the moving dislocations and contribute to additional hardening. In this regard, the solidification simulations for $G=10^7\ \mbox{K}/\mbox{m}$ were re-run, \textit{albeit} with two modifications. In one case, the backstress tensor, $\boldsymbol{\chi}$, was forced to be equal to zero, i.e., suppressing the contribution of GNDs to the directional residual stresses. In the second case, a uniform solute concentration was artificially imposed by setting the Cr concentration equal to the equilibrium concentration of $c_{Cr}^0=0.21$, i.e., suppressing solute segregation and its contribution to the GND development and backstress. Post-solidification, these simulations were cooled (at a cooling rate of $3\times10^5\ \mbox{K}/\mbox{s}$) and subsequently deformed in tension and compression along the BD as well as TD directions, respectively. Figure \ref{fig:figure_10}(a) shows the mechanical response in terms of the stress-strain curves and Figure \ref{fig:figure_10}(b) shows the predicted values of SD for these two cases, as compared with the predictions from a fully coupled PF-SGP model. These results clearly point out that the contribution of GND-induced backstress to the SD is higher than that due to solute segregation. Such an observation can be seen for the microstructrures loaded along both BD and TD. Modeling frameworks for predicting the anisotropic mechanical response of rapidly solidified microstructures therefore need to consider both these effects. In experiments, both the mechanisms are expected to act concurrently during rapid solidification, although the dominant mechanism may vary based on the microstructural factors and the boundary conditions imposed on the specimen.

We note that strain gradient crystal plasticity models (instead of $J_2$ plasticity models) could be better suited for predicting heterogeneous substructure development at the slip-system level during rapid solidification. However, these simulations become significantly computationally intensive when coupled with PF due to the fact that adaptive remeshing at the solidification front in our coupled simulations lead to over $300,000$ finite elements in the simulation domain during the solidification process. A representative solidification and cooling simulation takes $\approx$ 250 hours on 200 processors in parallel. A crystal plasticity simulation would take at least an order of magnitude longer time to run. Nonetheless, our $J_2$ SGP model has consideration for anisotropic elasticity and the crystallographic orientation based anisotropic factor, which are expected to influence to heterogeneous microstructure development and mechanical properties. As has been shown in Figures \ref{fig:figure_03} - \ref{fig:figure_10}, our model is able predict these heterogeneous microstructural attributes and anisotropic mechanical properties. We also note that a generalized plane strain was used in our 2D simulations. This implies that the microstructure is scalar along the third direction, $3$, and is a deviation from realistic rapid solidification. Presumably, the heterogeneity of microstructure development and mechanical properties would be further enhanced on consideration of 3D simulations. This will be explored in future work. Additionally, the inherent crystallographic texture of rapidly solidified microstructures has also been shown to be one of the factors contributing to the anisotropy in mechanical response \citep{geiger2016texture, jeon2019micro}, but in this work we have focused primarily on the role of microscale residual stresses on the TC asymmetry effect. The combined role of texture and microscale residual stresses on the TC asymmetry of AM microstructures may be explored in future work.

\section{Conclusions}

We propose a coupled phase field-strain gradient plasticity (PF-SGP) framework, to predict the TC asymmetry in rapidly solidified Fe-Cr alloys. The numerical framework, summarized in Equations \ref{phi_evolution}-\ref{backstress_calc}, accounts for multi-grain interaction effects, solute segregation, anisotropy of elastic and plastic deformation, GND density, directional backstress, local solute strengthening and thermally induced residual stress development in the microstructures during rapid solidification. The model was implemented in an open source finite element framework, MOOSE \citep{permann_moose_2020}, using a two-step scheme: firstly, the coupled PF-SGP framework was used to predict the microstructure evolution in terms of solute segregation, dislocation substructures and backstresses during rapid solidification of the liquid melt, under three different thermal gradients ($G=10^5\ \mbox{K}/\mbox{m}$, $10^6\ \mbox{K}/\mbox{m}$ and $10^7\ \mbox{K}/\mbox{m}$). Following this, only the SGP model was employed while cooling the microstructures (to RT) (with a cooling rate of $3\times10^3\ \mbox{K}/\mbox{s}$, $3\times10^4\ \mbox{K}/\mbox{s}$ and $3\times10^5\ \mbox{K}/\mbox{s}$), and subsequently deforming them in tension and compression along the build and the transverse directions separately. 

The coupled framework successfully captured the development of several microstructural features, such as Cr solute segregation, GND localizations and backstress evolution, during the rapid solidification of Fe-Cr alloys under different (imposed) thermal gradients. The significant model predictions can be summarized as follows: 

\begin{enumerate}
    \item Inter-dendritic regions or HAGBs accumulated large magnitudes of $\rho_{GND}$. Solute segregation introduced additional local lattice incompatibilities, further enhancing these $\rho_{GND}$ localizations. 
    \item A rise in $\rho_{GND}$ by $\sim 1-2$ orders of magnitude, and by an order of magnitude in $\bar{\chi}$ was observed per order of magnitude increase in $G$ (cf. Figures \ref{fig:figure_06}(a,b)).
    \item Post solidification and cooling, the room temperature mechanical response of these rapidly solidified microstructures, indicated that the predicted TC asymmetry, quantified in terms of the Strength Differential (SD), is in the range of reported experimental data (cf. Figure \ref{fig:figure_09}). Additionally, the SD values showed a rise with increase in the thermal gradient.
    \item The observed TC asymmetry could be explained in terms of the negative (average) backstress component along the transverse direction and a positive (average) backstress component along the build direction (cf. Figure \ref{fig:figure_08}).
    \item Simulations without solute segregation and those without backstress indicate that the thermal distortion induced GNDs (and hence backstresses) play a primary role in the observed TC asymmetry, although the TC contribution due to solute segregation is not negligible either (cf. Figure \ref{fig:figure_10}).
\end{enumerate}

The novelty of this modeling framework lies in its ability to predict the \textit{in situ} development of dislocation substructures and backstresses during rapid solidification that influence its post-solidification anisotropic mechanical properties. The modeling tools and analysis of model predictions in this study could be used to develop process-microstructure-mechanical property correlations in rapidly solidified microstructures and inform the appropriate choice of AM process parameters. The modeling framework will be extended in future work to simulate post-solidification heat treatments and their effects on the mechanical properties.

\appendix

\section{Appendix A: Strain Gradient Plasticity Model Predictions of Single Crystal Deformation}
\label{sec:appendix_A}

\begin{figure}[!htbp]
	\centering
	\includegraphics[width=0.9\textwidth]{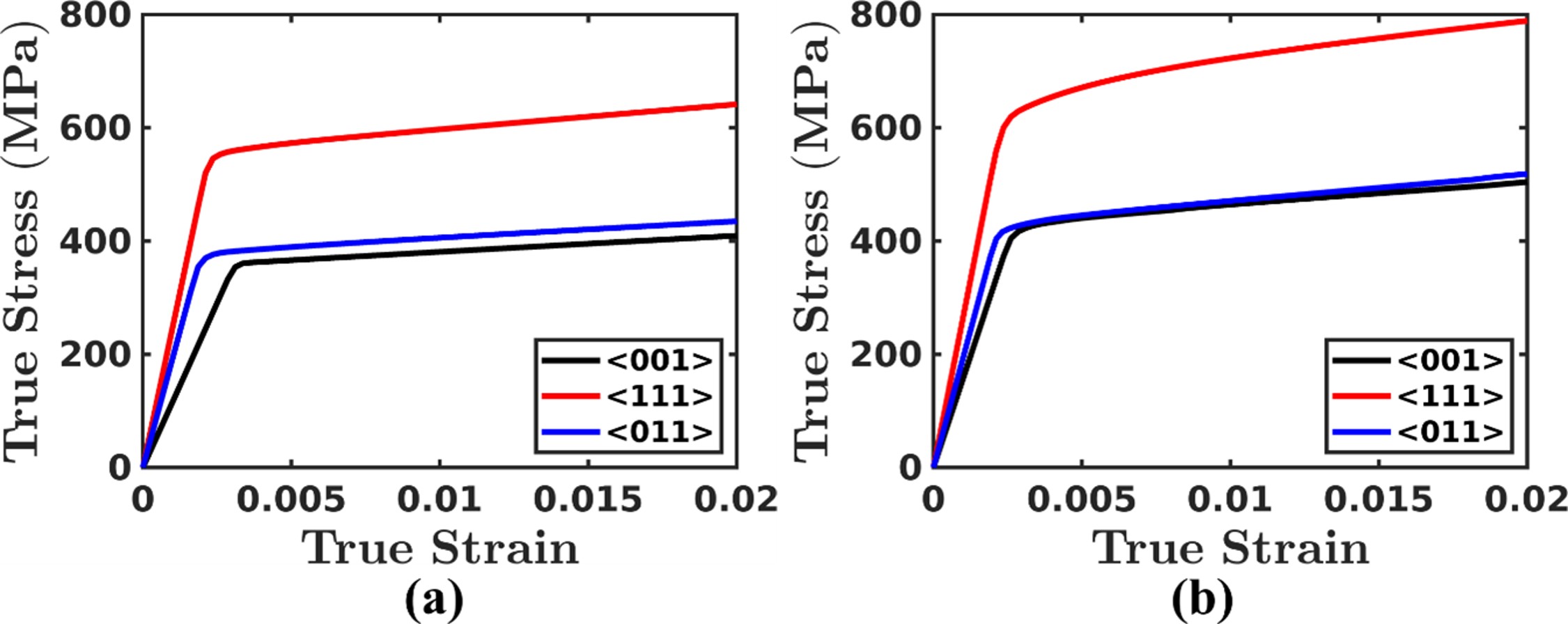}
	\hspace{1mm}
	\caption{ Model Predictions of mechanical response of $<001>$, $<011>$ and $<111>$ oriented single crystals obtained from (a) 3D and (b) 2D (plane strain) SGP simulations.}
	\label{fig:figure_A1}
\end{figure}

Figure \ref{fig:figure_A1} shows the uniaxial stress-strain behavior for single crystal simulations performed in 3D (Figure \ref{fig:figure_A1}(a)) and in 2D-plane strain (Figure \ref{fig:figure_A1}(b)), with crystals oriented along the $<001>$, $<011>$ and $<111>$ directions, respectively. A comparison of the predicted elastic modulus and yield strength with their corresponding analytical estimates has been provided in Table \ref{table_A1}. For a three dimensional domain, the analytical elastic modulus \citep{dieter1976mechanical} and yield strength \citep{patra_modeling_2023} is given by:

\begin{equation}
    \label{eqn:a_1}
    \frac{1}{E_{h k l}}=S_{11}-2\left[\left(S_{11}-S_{12}\right)-\frac{1}{2} S_{44}\right]\left(l^2 m^2+m^2 n^2+l^2 n^2\right)
\end{equation}

\begin{equation}
    \label{eqn:a_2}
    \bar{\sigma}_y \approx \bar{\chi}+a\left(\tau_0+k_{I H} G b \sqrt{\rho_{S S D}}\right)+a\left(s_t^0+k_{L N} \varepsilon_b ^ {\frac{4}{3}} c_{C r}^{\frac{2}{3}}\right)\left(1-\left(\frac{k T}{\Delta F_g} \log \left(\frac{\dot{\varepsilon}_0^p}{\bar{\varepsilon}^p}\right)\right)^{\frac{1}{q}}\right)^{\frac{1}{p}}
\end{equation}

where, $S_{ij}$ are the components of the compliance tensor, $\boldsymbol{S}$, and $l$,$m$,$n$ are the direction cosines for the crystal orientation under consideration. Equation \ref{eqn:a_2} is obtained by inverting the flow rule given in Equation \ref{eqn:equation_21} and the reader is referred to \cite{patra_modeling_2023} for further description. Note that the value of $\bar{\chi}$ has been directly added from our simulations. For the 2D simulations, we rotate the elastic tensor to a plane strain coordinate system \citep{shishvan2011plane}. The resulting plane strain elastic modulus $E^{*}$ can be analytically estimated as \citep{shishvan2011plane}:

\begin{equation}
    \label{eqn:a_3}
    E^*=\left[S_{11}^{\prime} \cos ^4 \phi+2\left(S_{12}^{\prime}+2 S_{66}^{\prime}\right) \cos ^2 \phi \sin ^2 \phi+S_{22}^{\prime} \sin ^4 \phi-\frac{\left(S_{13}^{\prime} \cos ^2 \phi+S_{12}^{\prime} \sin ^2 \phi\right)^2}{S_{33}^{\prime}}\right]^{-1}
\end{equation}

where, $S_{ij}^{'}$ denotes the components of the compliance tensor in the plane strain coordinate system, and $\phi$ is the angle between $X$ axis of the global and the plane strain coordinate system. Further, the methodology for estimating yield strength for a 2D domain is identical to that explained for its 3D counterpart.

Table \ref{table_A1} shows that the SGP predicted elastic moduli and yield stresses are in concurrence with the analytical values for all orientations. Generally speaking, the $<111>$ orientation is elastically stiffer \citep{brown_situ_2017} and also has a lower Schmid factor $(max(m^{\alpha}))$, in comparison to the $<011>$ and $<001>$ oriented single crystals. Similar observations have been noted for the 2D plane strain simulations, see Table \ref{table_A1}. In summary, these results indicate that the proposed SGP model demonstrates the capabilities to predict the crystal anisotropy induced deformations similar to a crystal plasticity model, at least for small strains, but at much lower computational costs in comparison to a crystal plasticity model.

\begin{table}[!htbp]
   \begin{center}
     \caption{\textcolor{black}{Comparison of the elastic modulus, $E$, and yield stress, $\sigma_y$, values between the single crystal SGP simulations and analytical calculations.}}
     \begin{tabular}{c c c c c}
     \hline
        \textcolor{black}{Simulation domain} & \textcolor{black}{Property} & \textcolor{black}{Crystal orientation} & \textcolor{black}{$J_{2}$ SGP predictions} & \textcolor{black}{Analytical calculations} \\ \hline
        \multirow{6}{1cm}{\textcolor{black}{3D}} & \multirow{3}{1cm}{\textcolor{black}{$E$ (GPa)}} & $\textcolor{black}{<001>}$ & $\textcolor{black}{116.51}$ & $\textcolor{black}{115.22}$ \\
         & & $\textcolor{black}{<011>}$ & $\textcolor{black}{192.81}$ & $\textcolor{black}{191.81}$ \\
         & & $\textcolor{black}{<111>}$ & $\textcolor{black}{247.67}$ & $\textcolor{black}{246.45}$ \\
          & \multirow{3}{1cm}{\textcolor{black}{$\sigma_{y}$ (MPa)}} & $\textcolor{black}{<001>}$ & $\textcolor{black}{303.51}$ & $\textcolor{black}{317.91}$ \\
         & & $\textcolor{black}{<011>}$ & $\textcolor{black}{320.65}$ & $\textcolor{black}{318.92}$ \\
         & & $\textcolor{black}{<111>}$ & $\textcolor{black}{528.81}$ & $\textcolor{black}{507.36}$ \\ \hline
         \multirow{6}{1cm}{\textcolor{black}{2D}} & \multirow{3}{1cm}{\textcolor{black}{$E$ (GPa)}} & $\textcolor{black}{<001>}$ & $\textcolor{black}{157.80}$ & $\textcolor{black}{156.93}$ \\
         & & $\textcolor{black}{<011>}$ & $\textcolor{black}{194.24}$ & $\textcolor{black}{191.83}$ \\
         & & $\textcolor{black}{<111>}$ & $\textcolor{black}{266.18}$ & $\textcolor{black}{263.10}$ \\
          & \multirow{3}{1cm}{\textcolor{black}{$\sigma_{y}$ (MPa)}} & $\textcolor{black}{<001>}$ & $\textcolor{black}{326.30}$ & $\textcolor{black}{326.10}$ \\
         & & $\textcolor{black}{<011>}$ & $\textcolor{black}{333.60}$ & $\textcolor{black}{326.34}$ \\
         & & $\textcolor{black}{<111>}$ & $\textcolor{black}{495.89}$ & $\textcolor{black}{516.32}$ \\ \hline
         
        \hline
     \end{tabular}
     \label{table_A1}
   \end{center}
\end{table}

\begin{figure}[!htbp]
	\centering
	\includegraphics[width=0.9\textwidth]{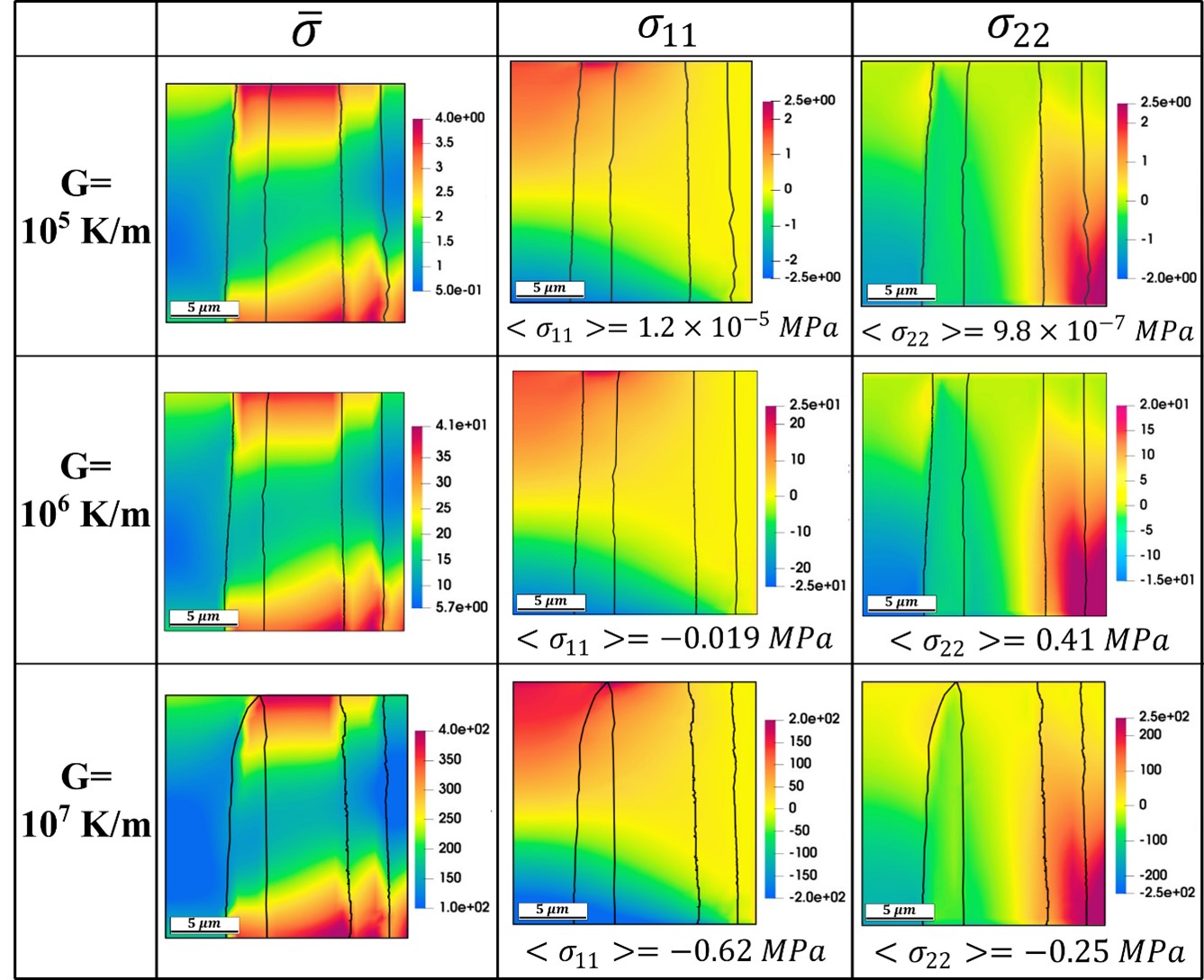}
	\hspace{1mm}
	\caption{ $\bar{\sigma}$, $\sigma_{11}$ and $\sigma_{22}$ contours in the fully solidified and cooled microstructures for the three different thermal gradients. Note that the scales (in MPa) are different in all three cases.}
	\label{fig:figure_B1}
\end{figure}

\section{Appendix B: Cauchy Stress Contours at the End of Solidification and Cooling}
\label{sec:appendix_B}

In the absence of any external loads, the residual Cauchy stresses within a simulation domain should self-equilibrate \citep{kapoor2018incorporating,bandyopadhyay2024initializing}, i.e., the net sum of the stress components should be equal to zero over the simulation domain. To verify this, the $\bar{\sigma}$, $\sigma_{11}$ and $\sigma_{22}$ contours were plotted for the fully solidified and cooled microstructures for the three different thermal gradients (prior to deformation) in Figure \ref{fig:figure_B1}. Further, the element area-weighted sum of the individual Cauchy stress components over the entire domain, i.e., $\left\langle\sigma_{i j}\right\rangle=\sum \frac{A_k}{A_{\text {total }}} \sigma_{i j}$, are also shown in Figure \ref{fig:figure_B1}. It can be seen that $\left|\left\langle\sigma_{i j}\right\rangle\right| \leq 1\ MPa$ for all three thermal gradients, thus verifying self-equilibration of residual stresses along the BD and TD. It can also be seen that the residual stresses are high in the regions where the backstress is high (see hotspots for $G=10^{7}$ K/m in Figure \ref{fig:figure_04} and the corresponding regions in Figure \ref{fig:figure_B1}, for example).

\section*{Acknowledgments}
The authors gratefully acknowledge funding received from the Department of Science and Technology (DST), India - Science and Engineering Research Board (SERB), India for this research under grant number: CRG/2020/000593. Namit Pai is also grateful for funding received from the Prime Minister’s Research Fellowship to support his doctoral studies. The support and the resources provided by PARAM Sanganak under the National Supercomputing Mission, Government of India, at the Indian Institute of Technology, Kanpur are gratefully acknowledged.

\bibliographystyle{apalike}  
\bibliography{cas-refs}

\end{document}